\documentclass[11pt]{article}
\usepackage[pdftex]{graphicx,color} 
\usepackage{jheppub}
\usepackage{amsmath}
\usepackage{amssymb}
\usepackage{comment}
\usepackage{multirow}
\usepackage{mathtools}
\usepackage{dsdshorthand}
\usepackage{mleftright}
\newcommand{\bea}{\begin{equation}\begin{aligned}}
\newcommand{\eea}[1]{\label{#1}\end{aligned}\end{equation}}
\newcommand{\beq}{\begin{equation}}
\newcommand{\eeq}{\end{equation}}

\newcommand   \zb  {\bar{z}}

\def\O{{\mathcal O}}

\newcommand\al{{\alpha}}

\newcommand   \ab  {\bar{\alpha}}

\usepackage{tikz}
\usetikzlibrary{arrows,calc,shapes,decorations.pathmorphing,decorations.markings}
\tikzset{
>=stealth',
help lines/.style={dashed, thick},
axis/.style={<->},
important line/.style={thick},
connection/.style={thick, dotted},
  cross/.style={
    cross out,
    draw=black, 
    minimum size=5pt, 
    inner sep=0pt,
    outer sep=0pt
  },
->-/.style={decoration={
  markings,
  mark=at position #1 with {\arrow{>}}},postaction={decorate}}
}

\title{AdS Virasoro-Shapiro amplitude with KK modes}

\author[a]{Giulia Fardelli,}
\author[b]{Tobias Hansen}
\author[b]{and Joao A. Silva}

\affiliation[a]{
\bigskip
Department of Physics and Astronomy\\
Uppsala University, Box 516, SE-751 20 Uppsala, Sweden
}

\affiliation[b]{
Mathematical Institute, University of Oxford,
Woodstock Road, Oxford, OX2 6GG, UK
\bigskip
}

\abstract{We determine the first curvature correction for the string amplitude of two supergravity states and two Kaluza-Klein modes on $\text{AdS}_5 \times \text{S}^5$, which is dual to the correlator $\langle \mathcal{O}_2 \mathcal{O}_2 \mathcal{O}_p \mathcal{O}_p \rangle$
of half-BPS operators in $\mathcal{N}=4$ SYM theory. The result has the form of an integral over the Riemann sphere as for the usual Virasoro-Shapiro amplitude, with the insertion of single-valued multiple polylogarithms of weight three.
The result fixes OPE data of single-trace operators in $\mathcal{N}=4$ SYM theory at strong coupling, including operators with non-zero $R$-charge and odd spin. We successfully check our results by comparing to data available from integrability, localisation and consistency with a $10d$ effective action.}

\emailAdd{giulia.fardelli@physics.uu.se, tobias.hansen@maths.ox.ac.uk, joao.silva@maths.ox.ac.uk}

\begin{document}
\maketitle

\section{Introduction}

Despite its prominent appearance in the AdS/CFT correspondence \cite{Maldacena:1997re}, the world-sheet theory of type IIB string theory on $\text{AdS}_5 \times \text{S}^5$
is still not fully known.
While the dual $\mathcal{N}=4$ supersymmetric Yang-Mills (SYM) theory is very well understood, 
making contact with the low energy effective action of strings in AdS requires studying $\mathcal{N}=4$ SYM at strong 't Hooft coupling $\lambda$.
The planar theory can be studied with integrability methods which give precise values for the dimensions of unprotected operators at finite 't Hooft coupling $\lambda$, see \cite{Gromov:2023hzc} for the latest results. However, at present, these methods are not developed enough to do the same for OPE coefficients.\footnote{See \cite{Caron-Huot:2022sdy, Cavaglia:2022qpg} for approaches that combine integrability and bootstrap.} 
In the series of papers \cite{Alday:2022uxp, Alday:2022xwz, Alday:2023jdk, Alday:2023mvu} a different approach was used to study the AdS Virasoro-Shapiro amplitude as an expansion in $\alpha'/R^2$ by combining properties of the putative world-sheet CFT and the dual CFT.
The result is an expression for the world-sheet integrand, an optimal target for future direct string theory computations.

In \cite{Alday:2022uxp, Alday:2022xwz, Alday:2023jdk, Alday:2023mvu} the correlator $\langle \mathcal{O}_2 \mathcal{O}_2 \mathcal{O}_2 \mathcal{O}_2 \rangle$ of four half-BPS operators of dimension two was studied. In this paper we study $\langle \mathcal{O}_2 \mathcal{O}_2 \mathcal{O}_p \mathcal{O}_p \rangle$ where $\mathcal{O}_p$ is a half-BPS operator of generic integer dimension $p$.\footnote{We do require that $p \ll \lambda^{\frac{1}{4}}$.} This enables us to study the OPE data of single trace operators of nonzero $R$-charge and odd spin, which appear in the OPE $\cO_2 \times \cO_p$.
Our main result is a formula for the first Anti-de Sitter (AdS) curvature correction $A^{(1)}(S, T)$ in $\langle \mathcal{O}_2 \mathcal{O}_2 \mathcal{O}_p \mathcal{O}_p \rangle$ (see section \ref{sec:Corr} and in particular formula \eqref{LEE} for the precise definition). It takes the form of a linear combination of terms, each of which is an integral over the Riemann sphere 
\bea
\int d^2 z |z|^{-2S-2}|1-z|^{-2T-2} G(S,T,z)\,,
\eea{IntegralMultiPolylogs}
where the integrand $G(S,T,z)$ is a sum of single-valued multiple polylogarithms of weight $3$, see section \ref{sec:WorldSheetCorr} for the precise formula. This can be contrasted with the flat space Virasoro-Shapiro amplitude $A^{(0)}(S, T)$ where the integrand is simply $1/(S+T)^2$
\bea
A^{(0)}(S, T) = \int d^2 z |z|^{-2S-2}|1-z|^{-2T-2} \frac{1}{(S+T)^2}\,.
\eea{FlatSpaceVirasoroIntro}
While the flat space Virasoro-Shapiro amplitude determines OPE data at leading order in the $\frac{1}{\sqrt{\lambda}}$ expansion, $A^{(1)}(S, T)$ determines the OPE data at subleading order.

The paper is organised in the following way. In section \ref{sec:Overview} we review basic facts about the correlator $\langle \mathcal{O}_2 \mathcal{O}_2 \mathcal{O}_p \mathcal{O}_p \rangle$, setting the stage for the computation of $A^{(1)}(S, T)$. Afterwards we compute $A^{(1)}(S, T)$ using two different algorithms. 

In section \ref{sec:Worldsheet} we apply the approach of \cite{Alday:2023mvu}, assuming that $A^{(1)}(S, T)$ is given by a linear combination of integrals of the form \eqref{IntegralMultiPolylogs}. We write an ansatz for the integrand in terms of 
single-valued multiple polylogarithms of $z$ and rational functions of $S$ and $T$.
This ansatz has a finite number of parameters, which are fixed by consistency of the residues of $A^{(1)}(S, T)$ with the OPE.

In section \ref{sec:Dispersion} we compute $A^{(1)}(S, T)$ by making use of dispersive sum rules, following \cite{Alday:2022xwz}. In this case we make an ansatz for the OPE data at subleading order in $\frac{1}{\sqrt{\lambda}}$ (which involves an infinite number of parameters). To fix these parameters we use locality (in the form of null constraints \cite{Caron-Huot:2020cmc, Caron-Huot:2021enk}). Besides this, we assume that the Wilson coefficients that enter in the strong coupling expansion are given by single-valued multiple zeta values. These two conditions fully fix all of the parameters.

Both methods have numerous internal consistency checks and they both agree with each other in the final answer.
In section \ref{sec:DataAndChecks} we extract OPE data and Wilson coefficients from $A^{(1)}(S, T)$ and compare them to the literature, namely results from localisation, integrability and the putative existence of an $AdS_5 \times S_5$ effective action, finding full agreement. Section \ref{sec:Conclusions} contains our final conclusions. We have numerous appendices with auxiliary results. In  Appendix \ref{app:ImprovedDisp} the fully crossing symmetric dispersion relation of \cite{Sinha:2020win, Gopakumar:2021dvg} is adapted to the present case of $\langle \mathcal{O}_2 \mathcal{O}_2 \mathcal{O}_p \mathcal{O}_p \rangle$ where only crossing symmetry between the $t$ and $u$ channels is present. This might be useful in other contexts.

\section{Basics}\label{sec:Overview}

\subsection{The correlator}\label{sec:Corr}

In this paper we consider correlation functions involving half-BPS operators $\mathcal{O}_p$ in planar $\mathcal{N}=4$ SYM, with SU$(N)$ gauge group,  as an expansion in inverse powers of the 't Hooft coupling $\lambda=g_{\mathrm{YM}}^2 N $. These half-BPS operators are Lorentz scalars transforming  in the $[0,p,0]$ representation of the SU$(4)_R$ symmetry group and they have protected conformal dimension $\Delta=p$. For $p=2$, this corresponds to the superprimary of the stress-tensor multiplet, while for $p>2$ they represent higher Kaluza-Klein supergravity excitations.
In terms of the $\mathcal{N}=4$ fundamental scalars, $\phi^I$, $I=1, \cdots, 6$, they  take the schematic form
\begin{align}
\mathcal{O}_{p}=y_{I_1} \cdots y_{I_p} \mathrm{tr}\left(\phi^{I_1} \cdots \phi^{I_p} \right)+\cdots
\end{align}
where we have contracted the SO$(6)_R \simeq \mathrm{SU}(4)_R$ indices with auxiliary null vectors $y_{I_i}$. The dots in the definition represent contributions from multi-trace operators, which are suppressed by inverse powers of the central charge $c=\frac{N^2-1}{4}$, and will not be relevant in the following.\footnote{The presence of these additional contributions, on top of the single-trace one, is necessary to make these operators dual to single-particle states in AdS and they can be fixed by requiring that each operator is orthogonal to any other multi-particle operator~\cite{Alday:2019nin, Aprile:2019rep,Aprile:2020uxk}. }

In this paper, we will focus on the four-point function involving two $\mathcal{O}_2$'s and two other identical $\mathcal{O}_p$'s. It can be written as~\cite{Dolan:2001tt, Dolan:2003hv,Dolan:2004iy,Caron-Huot:2018kta}
\bea
&\, \,  \langle \mathcal{O}_2(x_1, y_1) \mathcal{O}_2(x_2, y_2)\mathcal{O}_p(x_3, y_3)\mathcal{O}_p(x_4, y_4) \rangle=\frac{y_{12}^2 y_{34}^p}{(x_{12}^2)^2 (x_{34}^2)^p} \mathcal{G}_{\{22pp\}}(U,V; \alpha, \bar{\alpha})\, , \\
&\mathcal{G}_{\{22pp\}}(U,V; \alpha, \bar{\alpha})=\mathcal{G}^{\mathrm{free}}_{\{22pp\}}(U,V; \alpha, \bar{\alpha})+\frac{(z-\alpha)(z-\bar{\alpha})(\zb-\alpha)(\zb-\bar{\alpha})}{(z \zb)^2(\alpha\bar{\alpha})^2}\cT(U,V)\, , 
\eea{eq:22ppPosition}
where definitions of cross-ratios and generalisations can be found in Appendix~\ref{app:BasicsAppendix}.  The function $\mathcal{G}_{\{22pp\}}^{\mathrm{free}}$ represents the free theory contributions and can be evaluated by means of Wick contractions --- see~\eqref{eq:Gfree} for its explicit expression.   The dynamical information is encoded in the \textit{reduced correlator} $\cT(U,V)$, which  can be further split  as
\begin{align} \label{eq:TlongShort}
\cT(U,V)=\cT^{\mathrm{short}}(U,V)+\cT^{\mathrm{long}}(U,V) \, .
\end{align}
The first part receives contributions from protected short operators, while the second part contains the contributions of long operators. See Appendix~\ref{app:BasicsAppendix} for more details.  As we will soon see, the dimensions and OPE coefficients of these long operators get corrections at large $\lambda$. 

For our purposes, it is convenient to consider the Mellin transform  $M(s_1, s_2)$ of the reduced correlator
\bea
\mathcal{T}(U,V) &= \int_{-i \infty}^{+i \infty} \frac{ds_1 ds_2}{(4 \pi i)^2} \, U^{\frac{s_1}{2}+2+\frac{p}{3}} \, V^{\frac{s_2}{2}-1-\frac{p}{6}} \, \Gamma_{22pp}(s_1,s_2) \, M(s_1,s_2)\,, \\
\Gamma_{22pp}(s_1,s_2) &= \Gamma\left(\frac{6-p}{3}-\frac{s_1}{2}\right)  \Gamma\left(\frac{2p}{3}-\frac{s_1}{2}\right)  \Gamma\left(\frac{p}{6}+1-\frac{s_2}{2}\right)^2  \Gamma\left(\frac{p}{6}+1-\frac{s_3}{2}\right)^2, 
\eea{DefMellin}
where $s_1+s_2+s_3=0$.\footnote{These Mellin variables are related to the more standard $s,t,u$ by
$s_1 = s - \frac{2p}{3}$, $s_2 = t - \frac{2p}{3}$, $s_3 = u - \frac{2p}{3}$.} Crossing symmetry of the correlator~\eqref{eq:22ppPosition} under the exchange of points $1\leftrightarrow 2$ and $3 \leftrightarrow 4$, translates for the Mellin amplitude to
\begin{align}
M(s_1,s_2)=M(s_1,s_3)\, .
\end{align}

At leading order in $1/c$ and in an expansion around large $\lambda$, the Mellin amplitude is given by the supergravity result ($\lambda \to \infty$) plus a series of ``stringy'' corrections, described by polynomials in the Mellin variables and representing contact Witten diagrams with higher derivative vertices \cite{Alday:2018pdi, Binder:2019jwn}
\begin{align}
M(s_1, s_2) &= \frac{4p}{\Gamma(p-1)} \frac{1}{(s_1-\frac{6-2p}{3})(s_2-\frac{p}{3})(s_3-\frac{p}{3})}  \label{MellinLEE}\\
&+ \sum_{b=0}^{\infty} \sum_{a=b}^{\infty} \frac{\Gamma (a+b+p+4)}{\Gamma (b+1) \Gamma(p) \Gamma(p-1)} s_1^{a-b} (s_2 s_3)^b \lambda^{-\frac{3}{2}-\frac{a}{2}-\frac{b}{2}} \left( \xi_{a, b}^{(0)}(p) + \frac{1}{\sqrt{\lambda}} \xi_{a, b}^{(1)}(p) + \ldots \right).
\nonumber
\end{align}
The first term is the supergravity correlator and $\xi_{a, b}^{(k)}(p)$ are usually referred to as Wilson coefficients.

Following \cite{Alday:2022xwz} we also consider the transform
\bea
A(S, T) \equiv 2 \Gamma (p-1) \Gamma (p) \int_{-i \infty + \kappa}^{+i \infty + \kappa} \frac{d \alpha}{2 \pi i} e^{\alpha} \alpha^{-p-4} M\left(\frac{2 \sqrt{\lambda} S}{\alpha}, \frac{2 \sqrt{\lambda} T}{\alpha}\right)\,,
\eea{ScatteringAmplitude}
which is like the flat space limit \cite{Penedones:2010ue}, except that it applies to each layer $\xi_{a, b}^{(k)}(p)$, whereas the flat space limit considers only $\xi_{a, b}^{(0)}(p)$.
Applied to the low energy expansion \eqref{MellinLEE}, it implies
\bea
A(S, T) &= \frac{1}{\lambda^{\frac{3}{2}}} \sum\limits_{k=0}^\infty \frac{1}{\lambda^{k/2}}  A^{(k)}(S, T)\,,\\
A^{(k)}(S, T) &= \text{SUGRA}^{(k)} + \sum_{b=0}^{\infty} \sum_{a=b}^{\infty} \frac{2^{a+b+1}}{\Gamma (b+1)} S^{a-b} (TU)^b  \xi_{a, b}^{(k)}(p)\,. 
\eea{LEE}
The supergravity terms $\text{SUGRA}^{(k)}$ are given by
\begin{align}\label{SUGRA}
 \mathrm{SUGRA}^{(0)}=\frac{1}{S T U}\,, \  \mathrm{SUGRA}^{(1)}= -\frac{p}{6}\, \frac{p S^2+2 (p-3) T U}{(S T U)^2}\,, \ \ldots \,, \  \mathrm{SUGRA}^{(k>p)} = 0\,,
\end{align}
and the general expression can be found in \eqref{eq:sugraGeneric}.
The leading contribution is the Virasoro-Shapiro amplitude in flat space
\beq
A^{(0)}(S, T) = - \frac{ \Gamma \left(- S\right) \Gamma \left(-T\right) \Gamma \left(- U \right) }{\Gamma \left(S +1\right) \Gamma \left( T+1\right) \Gamma \left( U +1\right) }\,,
\eeq
which fixes the first layer of Wilson coefficients $\xi_{a, b}^{(0)}(p)$.

\subsection{Dispersion relation}

The ingredients to write a dispersion relation for the Mellin amplitude are its physical poles and residues and its behaviour at infinity.
As briefly discussed before, and more widely in Appendix~\ref{app:BasicsAppendix}, the $(12)$ OPE of $\cT(U,V)$ contains superconformal primaries that are singlets under SU$(4)_R$. We will indicate them by $\cO_s$: $\tau_s$ will label their twist and  $C_{s}$ the product of OPE coefficients $\langle \O_2 \O_2 \O_{s}  \rangle  \times \langle \O_p \O_p \O_{s} \rangle$.  In the Mellin amplitude, the exchange of these operators and their descendants results in an infinite sequence of poles at $s_1= \tau_s + 2 m -\frac{2}{3}p$, for $m \in \mathbb{N}_0$.
Similarly, the Mellin amplitude has also poles in $s_2$ and $s_3$, due to the exchange of superprimaries in the other OPE channels. We will denote them by $\cO_t$. They transform in the $[0, p-2, 0]$ representation of  $R$-symmetry, have twist  $\tau_t$  and we will use  $C_{t}$ to label the product of OPE coefficients $\langle \O_2 \O_p \O_t \rangle \times \langle \O_2 \O_p \O_t\rangle$.  The exact position of their poles is at  $s_2, s_3= \tau_t + 2 m- \frac{2}{3}p$, for $m \in \mathbb{N}_0$.

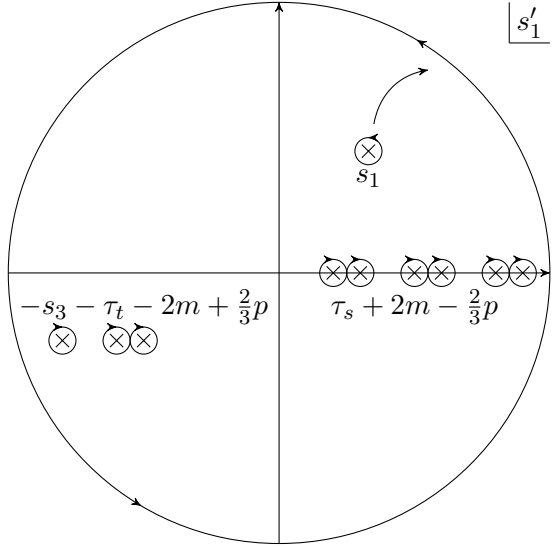
\begin{figure}
\centering
  \begin{tikzpicture}[scale=1.8]
    \coordinate (n) at (-0,2);
    \coordinate (e) at (2,0);
    \coordinate (w) at (-2,0);
    \coordinate (s) at (-0,-2);
    \coordinate (bp1) at (1,0);
    \coordinate (bp2) at (0,0);
    \draw[->] (w) --  (e) ;
    \draw[->] (s) --  (n) ;
    \node at (0.66,.9) [cross] {};  
    \draw[->-=.5] (0.66,0.8) arc (-90:270:0.1);
    \node at (0.66,0.7) [] {$s_1$};    
    \draw [->] (0.7,1.1) to [out=80,in=190] (1.1,1.5);    
    \node at (1.85,1.85) [] {$s_1'$};
    \draw[-] (2,1.7) -- (1.7,1.7);
    \draw[-] (1.7,1.7) -- (1.7,2);
    \draw[->-=.33] (2,0) arc (0:180:2);
    \draw[->-=.33] (-2,-0) arc (180:360:2);
    \node at (.4,0) [cross] {};
    \node at (.6,0) [cross] {};
    \node at (1.0,0) [cross] {};
    \node at (1.2,0) [cross] {};
    \node at (1.6,0) [cross] {};
    \node at (1.8,0) [cross] {};
    \draw[->-=.5] (0.4,-.1) arc (270:-90:0.1);
    \draw[->-=.5] (0.6,-.1) arc (270:-90:0.1); 
    \draw[->-=.5] (1.0,-.1) arc (270:-90:0.1); 
    \draw[->-=.5] (1.2,-.1) arc (270:-90:0.1); 
    \draw[->-=.5] (1.6,-.1) arc (270:-90:0.1); 
    \draw[->-=.5] (1.8,-.1) arc (270:-90:0.1); 
    \node at (-1.0,-0.5) [cross] {};
    \node at (-1.2,-0.5) [cross] {};
    \node at (-1.6,-0.5) [cross] {};
    \draw[->-=.5] (-1.0,-.6) arc (270:-90:0.1); 
    \draw[->-=.5] (-1.2,-.6) arc (270:-90:0.1); 
    \draw[->-=.5] (-1.6,-.6) arc (270:-90:0.1); 
    \node at (1,-0.25) [] {$\tau_s+2m - \frac{2}{3}p$};
    \node at (-1,-0.25) [] {$-s_3-\tau_t-2m +\frac{2}{3}p$}; 
  \end{tikzpicture}
\caption{Contour deformation involved in the dispersion relation. The integral around a generic point is equal to the sum over the physical poles plus a contribution from the arc at infinity. The contribution coming from the arc at infinity vanishes due to the bound on chaos.} \label{fig:contour_deformation}
\end{figure}

Due to the bound on chaos, \cite{Maldacena:2015waa} the planar Mellin amplitude is bounded in the Regge limit \cite{Penedones:2019tng}\footnote{The condition $\Re(s_2) \leq \frac{p}{3}$ comes from the fact that in the correlator $\langle \O_2 \O_2 \O_p \O_p \rangle$ the lowest twist exchanged in the $t$ channel has twist equal to $p$.}
\beq\label{bound chaos}
\lim_{s_1 \rightarrow \infty} M(s_1, s_2) \leq O(s_1^{-2})\,, \,\,\, \Re(s_2) \leq \frac{p}{3}\,.
\eeq
This allows us to write a dispersion relation for the Mellin amplitude $M(s_1, s_2)$.  To do that, we keep the variable $s_3$ fixed and rewrite the Mellin amplitude at a generic point as a contour integral in the $s_1$ plane. Afterwards, we deform the contour  as illustrated in figure \ref{fig:contour_deformation}.
The dispersion relation reads
\bea
M(s_1, s_2) &= \oint_{s_1} \frac{ds'}{2 \pi i} \frac{M(s', -s_3-s')}{s'-s_1}\\
 &= \sum_{\O_s} \sum_{m=0}^{\infty} C_{s} \frac{\mathcal{Q}_s(s_3; \tau_s, \ell, m) }{s_1- \tau_s-2m + \frac{2}{3}p} + \sum_{\O_t} \sum_{m=0}^{\infty} C_{t} \frac{\mathcal{Q}_t(s_3;  \tau_t, \ell, m) }{s_2- \tau_t-2m + \frac{2}{3}p} \,,
\eea{TwoChannelDisp}
where $\ell$ labels the spin of the exchanged operator. The kinematic functions $\mathcal{Q}_s(s_3;  \tau_s, \ell, m)$ and $\mathcal{Q}_t(s_3;  \tau_t, \ell, m)$ are related to Mack polynomials \cite{Mack:2009mi} and their expressions can be found in Appendix~\ref{app:Macks}.

Similarly to \eqref{eq:TlongShort}, $M(s_1, s_2)$ receives contributions from short and long multiplets.  When plugged in into the dispersion relation \eqref{TwoChannelDisp}, the information of protected operators\footnote{See Appendix~\ref{app:BasicsAppendix} for explicit OPE coefficients.} completely determines the supergravity amplitude.  Away from the strict $\lambda\to \infty$, single-trace long multiplets start  gaining corrections to their OPE data, thus producing the ``stringy" $\frac{1}{\sqrt{\lambda}}$ corrections in \eqref{MellinLEE}. We will comment more on that in the next section.\footnote{Double trace operators do not contribute to \eqref{TwoChannelDisp} at $\mathcal{O}(\frac{1}{c})$.}

\subsection{OPE data of stringy operators}

At strong coupling, the superprimaries belonging  to  long multiplets gain large anomalous dimensions.  Holographically these are related to  massive string states. In fact, at  leading order in $\frac{1}{\sqrt{\lambda}}$, their twist $\tau$  is given by $m R$, where $m$ is the mass of the string state and $R$ is the radius of AdS. In type IIB string theory, the string energy levels are $m^2 = 4 \frac{\delta}{\alpha'}$, with $\delta \in \mathbb{N}$.  As a consequence 
\begin{align}
\tau = 2 \sqrt{\delta} \, \lambda^{\frac{1}{4}}, \, \, \, \delta \in \mathbb{N}\,,
\end{align}
where recall  $\lambda= \frac{R^4}{\alpha'^2}$. Apart from $\delta$, the OPE data  of each stringy operator will depend  on $\lambda$, $p$, the spin $\ell$ and on other possible quantum numbers, which we will collectively denote by $\hat{r}$.  At leading order, multiple operators can share the same $\delta, \ell$ and transform in the same $R$-symmetry representations. This degeneracy (i.e.\ the range of $\hat{r}$) has been discussed in \cite{Alday:2023flc} and,  unfortunately, can not be resolved by just looking at four-point functions of half-BPS operators.  As a consequence, in our setup,  we will not be able to obtain OPE data for the unmixed operators, but just to compute averages over the quantum numbers $\hat{r}$ 
\begin{align}
\langle \ldots \rangle_{\delta,\ell} = \sum_{\hat{r}} \ldots \,.
\end{align}
When the exchanged operators have very large dimensions, as in this case, we can actually perform the sums  $\sum_{m=0}^{\infty}$ in \eqref{TwoChannelDisp}. 
It was shown in \cite{Alday:2022uxp} that this sum is dominated by $m \sim \tau^2$,
so we can set $m=x \tau^2$ and replace the sum with the integral
\beq
\sum_{m=0}^\infty \to \tau^2 \int_0^\infty dx\,.
\label{mtox}
\eeq
Some details on this can be found in Appendix \ref{app:LargeTwist}.
In this way we can completely fix the functional dependence of  the OPE data  on $\lambda$.  The twists go as
\bea
\tau_s(\delta, \ell, \hat{r}; \lambda, p) &= 2 \sqrt{\delta} \, \lambda^{\frac{1}{4}} + \tau_{s, 1}(\delta, \ell, \hat{r}; p)  +  \tau_{s, 2}(\delta, \ell, \hat{r}; p) \lambda^{-\frac{1}{4}} + \ldots \,,\\
\tau_t(\delta, \ell, \hat{r}; \lambda, p) &= 2 \sqrt{\delta} \, \lambda^{\frac{1}{4}} + \tau_{t, 1}(\delta, \ell, \hat{r}; p)  +  \tau_{t, 2}(\delta, \ell, \hat{r}; p) \lambda^{-\frac{1}{4}} + \ldots \,, 
\eea{twistsStringy}
and the OPE coefficients read
\bea
C_s(\delta, \ell, \hat{r}; \lambda, p) &=  \frac{\pi ^3 (-1)^p  \,\tau_s(\delta, \ell, \hat{r}; \lambda, p) ^{2 p+2} \, 2^{-2 \ell-2 p-2 \tau_s(\delta, \ell, \hat{r}; \lambda, p) -8}}{(\ell+1) \Gamma (p-1) \Gamma (p) \sin ^2\left(\frac{\pi  \tau_s(\delta, \ell, \hat{r}; \lambda, p) }{2}\right)}  f_s(\delta, \ell, \hat{r}; \lambda, p)\,,  \\
C_t(\delta, \ell, \hat{r}; \lambda, p) &=  \frac{\pi ^3 (-1)^{\ell} \tau_t(\delta, \ell, \hat{r}; \lambda, p) ^{2 p+2} 2^{-2 \ell-2 p-2 \tau_t(\delta, \ell, \hat{r}; \lambda, p) -8}}{(\ell+1) \Gamma (p-1) \Gamma (p) \sin ^2\left(\frac{\pi  p}{2}+\frac{\pi  \tau(\delta, \ell, \hat{r}; \lambda, p) }{2}\right)}  f_t(\delta, \ell, \hat{r}; \lambda, p)\,,  \\
f_s(\delta, \ell, \hat{r}; \lambda, p) &= f_{s, 0}(\delta, \ell, \hat{r}; p) +  f_{s, 1}(\delta, \ell, \hat{r}; p) \lambda^{-\frac{1}{4}} + f_{s, 2}(\delta, \ell, \hat{r}; p) \lambda^{-\frac{1}{2}} + \ldots  \,,   \\
 f_t(\delta, \ell, \hat{r}; \lambda, p) &= f_{t, 0}(\delta, \ell, \hat{r}; p) +  f_{t, 1}(\delta, \ell, \hat{r}; p) \lambda^{-\frac{1}{4}} + f_{t, 2}(\delta, \ell, \hat{r}; p) \lambda^{-\frac{1}{2}} + \ldots \,. 
\eea{OPEStringy}

\section{Worldsheet method}\label{sec:Worldsheet}

In this section we determine $A^{(1)}(S,T)$ using the method developed in \cite{Alday:2023jdk} for the correlator $\langle \mathcal{O}_2 \mathcal{O}_2 \mathcal{O}_2 \mathcal{O}_2 \rangle$.
We make an ansatz for the world-sheet integrand which has transcendentality three and contains single-valued multiple polylogs and single-valued multiple zeta values.
This was motivated in \cite{Alday:2023jdk, Alday:2023mvu} with a simple toy model of strings in AdS expanded around flat space.
The parameters of the ansatz are fixed by matching the residues with those predicted by the OPE.
We determine these residues directly by applying the Borel transform \eqref{ScatteringAmplitude} to the OPE poles of the Mellin amplitude in \eqref{TwoChannelDisp}.

\subsection{Poles and residues}
\label{sec:PolesResidues}

To discuss how the poles and residues of $A(S,T)$ in the $1/\sqrt{\lambda}$ expansion depend on OPE data, we follow Appendix C of \cite{Alday:2022xwz}.
For the Mellin amplitude we know that non-perturbatively all its poles originate from the OPE and have the form (we consider the poles in $s_1$ for concreteness)
\beq
M(s_1,s_2) \sim \frac{C_s \mathcal{Q}_s(s_3;\tau,\ell,m)}{s_1-\tau-2m+\frac23 p}\,,
\eeq
where $m$ labels conformal descendants.
We will see that each conformal family will become one particle in the flat space limit, so we have to consider the sum over descendants\footnote{Considering this sum over poles is justified by the dispersion relation \eqref{TwoChannelDisp}.}
\beq
M_{\tau,\ell}(s_1,s_2) = \sum_{m=0}^\infty
\frac{C_s \mathcal{Q}_s(s_3;\tau,\ell,m)}{s_1-\tau-2m+\frac23 p}\,.
\label{Mtauell}
\eeq
The transform \eqref{ScatteringAmplitude} of \eqref{Mtauell} is
\beq
A_{\tau,\ell}(S,T) = 
2 \Gamma (p-1) \Gamma (p) \int_{-i \infty + \kappa}^{+i \infty + \kappa} \frac{d \alpha}{2 \pi i} e^{\alpha} \alpha^{-p-4} 
\sum_{m=0}^\infty
\frac{C_s \mathcal{Q}_s(\frac{2 \sqrt{\lambda} U}{\alpha};\tau,\ell,m)}{\frac{2 \sqrt{\lambda} S}{\alpha}-\tau-2m+\frac23 p}\,.
\eeq
We exchange the integral and summation and for each $m$ pick the pole at\footnote{At this point we should explain why we do not pick up the pole at $\alpha=0$. The reason is the absence of a bulk point UV singularity of planar $\mathcal{N}=4$ SYM at finite 't Hooft coupling $\lambda$, as discussed in \cite{Maldacena:2015iua}. In the bulk point limit, we expect a divergence like $\frac{1}{(z-\bar{z})^{d-3}} = \frac{1}{z-\bar{z}}$. Translating that to Mellin space implies that the Mellin amplitude grows at most as $M(\xi S, \xi T) \sim \xi^{-p-4}$ in the limit of very large $\xi$ \cite{Penedones:2019tng}.}
\beq
\alpha = \alpha_* \equiv \frac{2 \sqrt{\lambda}S}{\tau+2m-\frac23 p}\,.
\eeq
Next we have to sum over $m$ as explained in \eqref{mtox}.
We arrive at
\beq
A_{\tau,\ell}(S,T) = 
- 2 \Gamma (p-1) \Gamma (p) \frac{\tau^2}{2\sqrt{\lambda} S}   \int_0^\infty dx \  e^{\alpha_*} \alpha_*^{-p-2}
C_s \mathcal{Q}_s\left(\tfrac{2 \sqrt{\lambda} U}{\alpha_*};\tau,\ell,x \tau^2\right)\,.
\eeq
We can now expand the integrand at large $\lambda$, using that $\tau \sim \lambda^{1/4}$.
The answer has the form
\beq
 -2 \Gamma (p-1) \Gamma (p) \frac{\tau^2}{2\sqrt{\lambda} S} e^{\alpha_*} \alpha_*^{-p-2} C_s
\mathcal{Q}_s\left(\tfrac{2 \sqrt{\lambda} U}{\alpha_*};\tau,\ell,x \tau^2\right)
= \frac{1}{\lambda^{\frac{3}{2}}} \frac{e^{-\frac{1}{4x} + \frac{\sqrt{\lambda}S}{\tau^2 x}}}{x^2} 
\sum_{i=0}^\infty \frac{1}{\lambda^{i/4}} P^{(i)} \left(\tfrac1x\right)\,,
\eeq
where $P^{(i)}\left(\tfrac1x\right)$ are polynomials in $\frac1x$ with increasing degree, for instance degree $\frac32 i$ for even $i$.
The leading term is given by
\beq
P^{(0)} \left(\tfrac1x\right) = \frac{\delta ^p S^{-p-3} f_{s;0}(\delta ,\ell) C_{\ell}^{(1)}\left(1+\frac{2 T}{S}\right)}{4 (\ell+1)}\,,
\eeq
where $C_{\ell}^{(1)}\left( x\right)$ is a Gegenbauer polynomial, as appropriate for partial waves in flat space.
We can now do the integrals in $x$ using
\beq
\int_0^\infty dx\, \frac{e^{-\frac{1}{4x} + \frac{\sqrt{\lambda}S}{\tau^2 x}}}{x^2} 
= - \frac{\frac{\tau^2}{\sqrt{\lambda}}}{S- \frac{\tau^2}{4\sqrt{\lambda}}}
= - \frac{4\delta}{S-\delta} + O\left( \lambda^{-\frac14}  \right)\,.
\label{x_integral}
\eeq
The effect of the polynomials $P^{(i)} \left(\tfrac1x\right)$ is that
each additional power of $\frac1x$ increases the order of the pole by one, as $\frac1x$
can be replaced by $\frac{\tau^2}{\sqrt{\lambda}}\partial_S$ acting on both sides of \eqref{x_integral}.

We first compute the poles and residues of $A_{\tau,\ell}(S,T)$ at order $\lambda^{-\frac{1}{4}}$. Since $A_{\tau,\ell}(S,T)$ should have no corrections at this order \eqref{LEE} we set the residues to zero, which fixes the OPE data at this order to the values given in \eqref{Solutionf1tau1} below.

At the next order $P^{(2)} \left(\tfrac1x\right)$ is a polynomial of degree 3 so that there are poles up to fourth order. The residues in terms of OPE data
are $p$-dependent generalisations of equation (C.22) of \cite{Alday:2022xwz}. We include them in an ancillary Mathematica file.
Crucially, the poles of order four and three depend only on OPE data that was fixed at lower orders.

\subsection{World-sheet correlator}\label{sec:WorldSheetCorr}

Next we make an ansatz for the world-sheet integral representation of our correlator that respects the crossing symmetry $A^{(1)}(S,T) = A^{(1)}(S,U)$
\beq
A^{(1)}(S,T) = B^{(1)}_1(S,T) + B^{(1)}_1(S,U) + B^{(1)}_1(U,T)
+ B^{(1)}_2(S,T) + B^{(1)}_2(S,U)\,,
\label{A_to_B}
\eeq
where
\begin{equation}
B^{(1)}_i (S,T)=\int d^2 z |z|^{-2S-2}|1-z|^{-2T-2} G^{(1)}_i(S,T,z)\,, \quad i=1,2\,,
\label{B}
\end{equation}
and crossing symmetry implies
\beq
B^{(1)}_1 (U,T) = B^{(1)}_1 (T,U)
\quad \Leftrightarrow \quad
G^{(1)}_1(S,T,z) = G^{(1)}_1(T,S,1-z)\,.
\label{G1_symmetry}
\eeq
Note that there is no symmetry constraint on $B^{(1)}_2 (S,T)$ and the only reason to include the terms $B^{(1)}_1(S,T)$ and $B^{(1)}_1(S,U)$ in the ansatz is to make it easier to compare to the fully crossing symmetric case $p=2$, since in this case it is easy to see that our ansatz implies
\beq
B^{(1)}_2 (S,T) =0\,, \quad p=2\,.
\eeq
We also impose symmetry under the exchange $z \leftrightarrow \bar{z}$ and define the following combinations of single-valued multiple polylogarithms\footnote{See \cite{Brown:2004ugm,Dixon:2012yy} and also \cite{Alday:2023jdk, Alday:2023mvu} for details on these functions in this context.}
\bea
\mathcal{L}^s_{w}(z) ={}& \mathcal{L}_{w}(z) + \mathcal{L}_{w}(1-z)+\mathcal{L}_{w}(\bar{z}) + \mathcal{L}_{w}(1-\bar{z})\,,\\
\mathcal{L}^a_{w}(z) ={}& \mathcal{L}_{w}(z) - \mathcal{L}_{w}(1-z)+\mathcal{L}_{w}(\bar{z}) - \mathcal{L}_{w}(1-\bar{z})\,.
\eea{L_symmetrised}
We choose the following weight 3 basis \cite{Alday:2023mvu}
\bea
\mathcal{L}^{(1)s} ={}& \Big(
\mathcal{L}^s_{000}(z),
\mathcal{L}^s_{001}(z),
\mathcal{L}^s_{010}(z),
\zeta(3)
\Big)\,,\\
\mathcal{L}^{(1)a} ={}& \Big(
\mathcal{L}^a_{000}(z),
\mathcal{L}^a_{001}(z),
\mathcal{L}^a_{010}(z)
\Big)\,.
\eea{k1_L_basis}
Our ansatz for the worldsheet correlator is then
\bea
G^{(1)}_i(S,T,z) &= \sum\limits_{u=1}^4 r^{(1)s}_{i,u} (S,T) \mathcal{L}^{(1)s}_u +\sum\limits_{u=1}^3 r^{(1)a}_{i,u} (S,T) \mathcal{L}^{(1)a}_u \,,
\eea{G_ansatz}
where
$r^{(1)s/a}_{i,u} (S,T)$ are rational functions of homogeneity 0 and we assume the denominator to be $U^2$. Furthermore $r^{(1)s/a}_{1,u} (S,T)$ have symmetry properties according to \eqref{G1_symmetry}
\beq
r^{(1)s}_{1,u} (S,T) = \frac{f_{1,u}(p) (S+T)^2 + f_{2,u}(p) S T}{(S+T)^2}\,, \qquad
r^{(1)a}_{1,u} (S,T) = f_{3,u}(p) \frac{S-T}{S+T}\,,
\eeq
whereas $r^{(1)s/a}_{2,u} (S,T)$ are of the form
\beq
\frac{f_{4,u}(p) (S+T)^2 + f_{5,u}(p) S T+ f_{6,u}(p) (S^2-T^2)}{(S+T)^2}\,.
\eeq
In total our ansatz depends on 32 functions of $p$.

We can now compute the residues of our ansatz at $S=1,2,\ldots$ and  $T=1,2,\ldots$ and match them with those computed in section \ref{sec:PolesResidues} in terms of OPE data. We further compute the residues at $S=0$ and $T=0$ and match them to the supergravity term \eqref{SUGRA}.
As discussed in \cite{Alday:2023mvu} the functions $G^{(1)}_i(S,T,z)$ have certain ambiguities, i.e.\ terms that we can add to $G^{(1)}_i(S,T,z)$ without changing $A^{(1)}(S,T)$.
With the ansatz discussed here, there are 6 such ambiguities, so that our final answer has the form
\beq
G^{(1)}_i(S,T,z) = \sum\limits_{u=1}^4 r^{(1)s}_{i,u} \mathcal{L}^{(1)s}_u +\sum\limits_{u=1}^3 r^{(1)a}_{i,u} \mathcal{L}^{(1)a}_u
+ \sum_{j=1}^{6} a_j \left( \sum\limits_{u=1}^4 \hat{r}^{(1)s}_{i,ju} \mathcal{L}^{(1)s}_u +\sum\limits_{u=1}^3 \hat{r}^{(1)a}_{i,ju} \mathcal{L}^{(1)a}_u  \right) \,,
\label{worldsheet_correlator}
\eeq
with
\bea
r^{(1)s}_1 &= \frac{1}{24} \left(-p^2,2 (p-2) p,p^2-2 p-6,48\right)\,,\\
r^{(1)a}_1 &= \frac{p^2(S-T)}{24(S+T)} \left(-1,2,1 \right)\,,\\
r^{(1)s}_2 &= \frac{p(p-2)}{24(S+T)} \left(3 S,-2 (2 S+T),-2 S-T,0 \right)\,,\\
r^{(1)a}_2 &= \frac{p(p-2)}{24(S+T)} \left(3 S,-2 (2 S-T),-2 S+T \right)\,.
\eea{r_result}
The coefficients $a_j$ are unfixed and multiply the ambiguities, which are given in appendix \ref{app:ambiguities}.

\section{Dispersive sum rules method}\label{sec:Dispersion}
In this section we compute $A^{(1)} (S,T)$ with a different method, following \cite{Alday:2022xwz}.
We apply the low energy expansion \eqref{MellinLEE} to a dispersion relation to obtain dispersive sum rules, which relate Wilson coefficients to OPE data.
We then make an ansatz in terms of nested sums for certain combinations of OPE data, and fix the parameters of the ansatz with locality and single-valuedness constraints for the Wilson coefficients.
 
In~\eqref{TwoChannelDisp} we have seen a way to write $M(s_1, s_2)$ dispersively in terms of two channels.   Unfortunately, written in this way,  the symmetry under the exchange of $s_2$ and $s_3$ is obscured. Therefore, to make it manifest, in this section we implement a different dispersion relation, where, instead of  $s_3$, we keep fixed a new parameter $r=\frac{s_2 s_3}{s_1}$.  With this change of coordinates,  $s_2$ and $s_3$ are now treated on the same footing, thus leading to a more symmetric dispersion relation and eventually to simpler sum rules.\footnote{In spirit, this is very similar to the crossing symmetric dispersion relations of \cite{Sinha:2020win, Gopakumar:2021dvg} for the fully crossing symmetric case.} The details on how this dispersion relation is derived are deferred to Appendix \ref{app:ImprovedDisp}. In the following we simply report the associated sum rules and the corresponding results for the Wilson coefficients and OPE data.

\subsection{Flat space OPE data}\label{secFlatSpaceOPEdata}
The first  of the dispersive sum rules relating Wilson coefficients and OPE data reads
\begin{align}\label{flatSpaceEqs}
\xi_{a, b}^{(0)}(p) &= \sum_{\delta=1}^{\infty} \sum_{q=0}^b \frac{c_{s,a,b,q}^{(0)} F_{s; q}^{(0)}(\delta)+c_{t,a,b,q}^{(0)} F_{t; q}^{(0)}(\delta)}{\delta^{3+a+b}}\,,
\end{align}
where 
\bea
 F_{s; q}^{(0)}(\delta) &= \frac{4^q}{\Gamma(2q+2)} \sum_{\ell=0, 2}^{2\delta-2} \langle f_{s, 0}\rangle_{\delta, \ell} (\ell-q+1)_q (2+\ell)_q\,, \\
 F_{t; q}^{(0)}(\delta) &= \frac{4^q}{\Gamma(2q+2)} \sum_{\ell=0, 1}^{2\delta-2} \langle f_{t, 0}\rangle_{\delta, \ell} (\ell-q+1)_q (2+\ell)_q\,,
\eea{defF0}
and
\bea
c_{s,a,b,q}^{(0)}&=\frac{(-1)^q q  2^{-a-b-1} \Gamma (2 b-q)}{\Gamma (b-q+1)}, ~~~ q>0\,, \\
c_{s,a,b,0}^{(0)}&=2^{-a-1} \delta_{b,0} , ~~~ q=0\,, \\
c_{t,a,b,q}^{(0)}&=\frac{(-1)^{a+1+q} \Gamma (a) \Gamma (b+1) 2^{-a-b-1} (-a-b+q) \, }{ \Gamma (b-q+1) \Gamma (a-b+q+1)} \\
&\times _3F_2(q,q-b,-a-b+q+1;-a-b+q,a-b+q+1;1) , ~~~ q>0\,,  \\
c_{t,a,b,0}^{(0)}&=\frac{(-2)^{-a-b-1} (a+b) \Gamma (b-a)}{\Gamma (1-a)}, ~~~ q=0\,.
\eea{cs0}
The Wilson coefficients $\xi_{a, b}^{(0)}(p)$ can be read off from the flat space Virasoro-Shapiro amplitude \big(see (\ref{FlatSpace1}) and (\ref{FlatSpace2})\big). Thus we can systematically analyse the equations (\ref{flatSpaceEqs}), solving for the functions $F_{s; q}^{(0)}(\delta)$ and $F_{t; q}^{(0)}(\delta)$. As an example, we work out the $b=0$ case in appendix \ref{app:bEquals0}. After studying many cases, the general conclusion we extract is that
\begin{align}\label{solF01}
F_{s; q}^{(0)}(\delta) = F_{t; q}^{(0)}(\delta) = F_{q}^{(0)}(\delta) , \ \forall q \in \mathbb{N}_0\,,
\quad \Leftrightarrow \quad
\langle f_{s, 0}\rangle_{\delta, \ell}
= \langle f_{t, 0}\rangle_{\delta, \ell}
= \langle f_{0}\rangle_{\delta, \ell}\,,
\end{align} 
where $F_{q}^{(0)}(\delta)$ and $\langle f_{0}\rangle_{\delta, \ell}$
are the same as in the $\langle \cO_{2} \cO_{2} \cO_{2} \cO_{2} \rangle$ case \citep{Alday:2022xwz}, i.e.\ 
\begin{align}\label{solF02}
F_{q}^{(0)}(\delta)= \sum\limits_{d=\lfloor \frac{q+1}{2} \rfloor}^q
\sum\limits_{\substack{s_1,\ldots,s_d \in \{1,2\}\\s_1+\ldots+s_d=q}}
2^{\sum_i \delta_{s_i,1}} \delta^q Z_{s_1,\ldots,s_d} (\delta-1)\, ,
\end{align}
where $Z_{s_1,\ldots,s_d} (\delta-1)$ is an Euler-Zagier sum defined by
\begin{align}
Z_{s_1, s_2, s_3, \ldots} (N) = \sum_{n=1}^N \frac{Z_{s_2, s_3, \ldots} (n-1)}{n^{s_1}}\,,\qquad
Z(N)=1\,,\qquad
 Z_{s_1, s_2, s_3, \ldots} (0) = 0\,.
\end{align}
In particular we find that $\langle f_{t, 0}\rangle_{\delta, \ell} =0$, if $\ell$ is odd. Also, we find that any $\langle f_{s, 0}\rangle_{\delta, \ell}$ does not depend on $p$, i.e.\ at leading order in $\lambda$ the full $p$-dependence of the OPE coefficients is in the prefactor in (\ref{OPEStringy}).
All this is in agreement with the more general result of \cite{Alday:2023flc} for the product of leading OPE coefficients $\langle \O_{p_1} \O_{p_2} \O_{\Delta,\ell,[0,n,0]}  \rangle  \times \langle \O_{p_3} \O_{p_4} \O_{\Delta,\ell,[0,n,0]} \rangle$.

At the next order we impose that there are no $\lambda^{-\frac{1}{4}}$ corrections to the flat space Wilson coefficients. From the string theory perspective, this would correspond to fractional powers of $\alpha'/R^2$, and we assume that such powers are absent. This leads to the equation
\bea\
0 ={}& \sum_{\delta=1}^{\infty} \sum_{q=0}^b \frac{c_{s,a,b,q}^{(0)}\big( F_{s; q}^{(1)}(\delta) - (3+a+b) 
T_{s; q}^{(1)}(\delta) \big)}{\delta^{\frac{7}{2}+a+b}}  \\
&+  \sum_{\delta=1}^{\infty} \sum_{q=0}^b \frac{c_{t,a,b,q}^{(0)}\big( F_{t; q}^{(1)}(\delta) - (3+a+b) 
T_{t; q}^{(1)}(\delta) \big)}{\delta^{\frac{7}{2}+a+b}} \,,
\eea{Eqsf1tau1}
where 
\begin{align}
F_{s; q}^{(1)}(\delta) &= \frac{4^q}{\Gamma(2q+2)} \sum_{\ell=0, 2}^{2\delta-2} (\ell-q+1)_q (2+\ell)_q \left( \sqrt{\delta} \langle f_{s, 1}\rangle_{\delta, \ell} - \langle f_{s, 0} \rangle_{\delta, \ell}\left(\ell p+\ell+2 p+\frac{7}{4}\right) \right) , \nonumber \\
T_{s; q}^{(1)}(\delta) &= \frac{4^q}{\Gamma(2q+2)} \sum_{\ell=0, 2}^{2\delta-2} \langle f_{s, 0}\rangle_{\delta, \ell} (\ell-q+1)_q (2+\ell)_q ( \tau_{s, 1} (\delta, \ell)  + \ell +2 ),\nonumber  \\
F_{t; q}^{(1)}(\delta) &= \frac{4^q}{\Gamma(2q+2)} \sum_{\ell=0, 1}^{2\delta-2} (\ell-q+1)_q (2+\ell)_q \left( \sqrt{\delta} \langle f_{t, 1}\rangle_{\delta, \ell}  - \langle f_{t, 0}\rangle_{\delta, \ell} \left( \ell p+\ell+\frac{p^2}{2}+\frac{15}{4}  \right) \right) , \nonumber \\
T_{t; q}^{(1)}(\delta) &= \frac{4^q}{\Gamma(2q+2)} \sum_{\ell=0, 1}^{2\delta-2} \langle f_{t, 0}\rangle_{\delta, \ell} (\ell-q+1)_q (2+\ell)_q ( \tau_{t, 1} (\delta, \ell) + \ell +2 ).
\label{t1f1}
\end{align}
The solution to these equations is\footnote{We assumed in this analysis that all operators acquire the same finite shift, as argued in section $6.1.1$ of \cite{Gromov:2023hzc}.}
\bea
\tau_{s, 1} (\delta, \ell)  &= \tau_{t, 1} (\delta, \ell)  = - 2 - \ell\,, \\
\langle f_{s, 1}\rangle_{\delta, \ell} &= \frac{ \langle f_{s, 0}\rangle_{\delta, \ell}}{\sqrt{\delta}} \left(\ell p+\ell+2 p+\frac{7}{4}\right) \,, \\
\langle f_{t, 1}\rangle_{\delta, \ell} &= \frac{ \langle f_{t, 0}\rangle_{\delta, \ell}}{\sqrt{\delta}} \left(\ell p+\ell+\frac{p^2}{2}+\frac{15}{4} \right) \,.
\eea{Solutionf1tau1}

\subsection{First curvature correction}

At the next order we find the dispersive sum rule
\bea
\xi^{(1)}_{a, b}(p) ={}&   \sum_{\delta=1}^{\infty} \sum_{q=0}^b \frac{c_{s,a,b,q}^{(0)}\big( F_{s; q}^{(2)}(\delta) - (3+a+b) 
T_{s; q}^{(2)}(\delta) \big) 
+    c_{s,a,b,q}^{(2,0)} F_{s; q}^{(0)}(\delta)+ c_{s,a,b,q}^{(2,1)} F_{s; q+1}^{(0)}(\delta)}{\delta^{4+a+b}}\\
&+ \sum_{\delta=1}^{\infty} \sum_{q=0}^b \frac{c_{t,a,b,q}^{(0)}\big( F_{t; q}^{(2)}(\delta) - (3+a+b) 
T_{t; q}^{(2)}(\delta) \big)  
+    c_{t,a,b,q}^{(2,0)}F_{t; q}^{(0)}(\delta)+ c_{t,a,b,q}^{(2,1)}F_{t; q+1}^{(0)}(\delta)  }{\delta^{4+a+b}}\,,
\eea{EquationSecondOrder}
where 
\begin{align}
F_{s; q}^{(2)}(\delta) &= \frac{4^q}{\Gamma(2q+2)} \sum_{\ell=0, 1}^{2\delta-2} (\ell-q+1)_q (2+\ell)_q \left( \delta  \langle f_{s, 2}\rangle_{\delta, \ell} -\frac{1}{4} (4 p^2+9 p+5) \ell  \langle f_{s, 0}\rangle_{\delta, \ell} \right) , \nonumber\\
T_{s; q}^{(2)}(\delta) &= \frac{4^q}{\Gamma(2q+2)} \sum_{\ell=0, 1}^{2\delta-2}  \sqrt{\delta}(\ell-q+1)_q (2+\ell)_q  \langle f_{s, 0}  \tau_{s, 2} \rangle_{\delta, \ell} , \nonumber\\
F_{t; q}^{(2)}(\delta) &= \frac{4^q}{\Gamma(2q+2)} \sum_{\ell=0, 1}^{2\delta-2} (\ell-q+1)_q (2+\ell)_q \left( \delta \langle f_{t, 2}\rangle_{\delta, \ell} - \frac{1}{4} \left(2 p^3-2 p^2+9 p+13\right) \ell  \langle f_{t, 0}   \rangle_{\delta, \ell} \right) , \nonumber\\
T_{t; q}^{(2)}(\delta) &= \frac{4^q}{\Gamma(2q+2)} \sum_{\ell=0, 1}^{2\delta-2}  \sqrt{\delta} (\ell-q+1)_q (2+\ell)_q \langle f_{t, 0}  \tau_{t, 2} \rangle_{\delta, \ell},
\label{T2F2}
\end{align}
and we wrote $c_{s,a,b,q}^{(2,0)}$, $c_{s,a,b,q}^{(2,1)}$, $c_{t,a,b,q}^{(2,0)}$ and $c_{t,a,b,q}^{(2,1)}$ in (\ref{dump1}). The OPE data (\ref{T2F2}) is truly sensitive to AdS curvature corrections, since it depends on Wilson coefficients $\xi^{(1)}_{a, b}(p)$ which are not fixed by the flat space Virasoro-Shapiro amplitude. The Wilson coefficients $\xi^{(1)}_{a, b}(p)$ are in general unknown.

The situation regarding the sum rules (\ref{EquationSecondOrder}) is now different from the one we encountered in section \ref{secFlatSpaceOPEdata}. In section \ref{secFlatSpaceOPEdata} the LHS of the sum rules was known and the Wilson coefficients were an input into the dispersive sum rules and the OPE data was the output. 
To make progress we will proceed along the same lines of \cite{Alday:2022xwz}. Fixing $b$, we make a general ansatz for the OPE data that enters (\ref{EquationSecondOrder}), as a linear combination of Euler-Zagier sums involving a few undetermined functions of $p$. Using (\ref{EquationSecondOrder}) this determines the Wilson coefficients in terms of those functions. Afterwards, we impose physical conditions on the Wilson coefficients (namely single-valuedness) and these will completely fix all of these functions.

Let us see how this comes about in practice. In complete analogy with the $p=2$ case, we first assume that $T_{s; q}^{(2)}(\delta)$ and $T_{t; q}^{(2)}(\delta)$ are linear combinations of Euler-Zagier sums with maximal depth $q+1$ and maximal weight $q+2$ and that $F_{s; q}^{(2)}(\delta)$ and $F_{t; q}^{(2)}(\delta)$ are linear combinations of Euler-Zagier sums with maximal depth $q+1$ and maximal weight $q+3$:
\bea
T_{s; q}^{(2)}(\delta)
={}& A^1_{q+2,q+1} , \\
T_{t; q}^{(2)}(\delta)
={}& A^2_{q+2,q+1} ,\\
F_{s; q}^{(2)}(\delta)
={}& A^3_{q+3,q+1} + \delta^3 \zeta(3) A^4_{q,q} , \\
F_{t; q}^{(2)}(\delta)
={}& A^5_{q+3,q+1} + \delta^3 \zeta(3) A^6_{q,q}   ,
\eea{T2F2def}
for $q \geq 1$ where 
\beq
A^i_{w_\text{max}, d_\text{max}} = \sum\limits_{w=1}^{w_\text{max}} \sum\limits_{d=1}^{d_\text{max}}
\sum\limits_{\substack{s_1,\ldots,s_d \in \mathbb{N}\\s_1+\ldots+s_d=w}}
c^i_{s_1,\ldots,s_d}(p) \delta^w Z_{s_1,\ldots,s_d} (\delta-1)\,.
\label{ansatz_A}
\eeq
$c^i_{s_1,\ldots,s_d}(p)$ are the functions of $p$ that are fixed by our procedure. For  $q=0$, see formula (\ref{qEquals0Ansatz}).

Second, we impose that the strong coupling expansion cannot contain negative powers of the Mellin variables apart from the supergravity term. This simple condition is nontrivial from the point of view of equations (\ref{EquationSecondOrder}) and leads to
\begin{align}\label{CrossingCondition}
\xi_{a, b}^{(1)}(p)=0, \,\, \text{for} \, \, a=0, \ldots, b-1, \forall \, b \in \mathbb{N} .
\end{align}
Plugging (\ref{EquationSecondOrder}) into (\ref{CrossingCondition}) leads to constraints on the OPE data.

Third, we impose that $\xi_{a, b}^{(1)}(p)$ is in the ring of single-valued multiple zeta values. For a practical introduction to this topic, see section $3.1$ of \cite{Alday:2022xwz}. We use the MAPLE program \textit{HyperlogProcedures} \cite{HyperlogProcedures} to do so. In practice, for any fixed $b$ it is enough to impose single-valuedness for a few values of $a$ to fix all of the undetermined functions. Afterwards, we check that the resulting expression for $\xi_{a,b}^{(1)}(p)$ is single-valued for other values of $a$ that we did not use before.\footnote{For example, for $b=0$, we impose single-valuedness for $a=0, \ldots, 6$ and checked it for $a=7, \ldots, 10$. For $b=1$ we imposed it for $a=1, \ldots, 8$ and checked it for $a=9, 10, 11$. For $b=2$ we imposed it for $a=2, \ldots, 8$ and checked it for $a=9, 10, 11$.}

In this manner we solved (\ref{EquationSecondOrder}) for $b=0, \dots, 5$. In appendix \ref{app:ExampleMaple} we write all the details on how to implement this for the $b=0$ case. From these cases, we extract the following general conclusions
\bea
T_{s; q}^{(2)}(\delta) &=  T_{q}^{(2)}(\delta)\,, \\
T_{t; q}^{(2)}(\delta) &=  T_{q}^{(2)}(\delta) + \frac{p^2-4}{4}  F_{q}^{(0)}(\delta)\,,
\eea{taus}
where $ T_{q}^{(2)}(\delta) $ is the value obtained in the $\langle \cO_2\cO_2\cO_2\cO_2 \rangle$ case \cite{Alday:2022xwz} for $T_{s; q}^{(2)}(\delta)$. 

As to the OPE coefficients, they are determined by
\bea
F_{s; q}^{(2)}(\delta) &= F_{q}^{(2)}(\delta) + (p-2) \big( g_1(q, p) F_{q}^{(0)}(\delta) + g_2(q, p) F_{q+1}^{(0)}(\delta)  \big)\,, \\
F_{t; q}^{(2)}(\delta) &= F_{q}^{(2)}(\delta) + (p-2) \big( g_3(q, p) F_{q}^{(0)}(\delta) + g_4(q, p) F_{q+1}^{(0)}(\delta)  \big)\,,
\eea{f2s}
where $F_{q}^{(2)}(\delta)$ is the value obtained in the $\langle \cO_2\cO_2\cO_2\cO_2 \rangle$ case \cite{Alday:2022xwz} for $F_{s; q}^{(2)}(\delta)$ and
\bea
g_1(q, p) &= -\frac{p^2}{3}+\frac{1}{6} p \left(3 q^2+3 q+5\right)+2 q^2+3 q+6 , \\
g_2(q, p) &= \frac{1}{2} (q+1) (p q+p+4 q+5), \\
g_3(q, p) &= \frac{1}{24} \left(3 p^3+4 p^2+p \left(12 q^2+18 q+41\right)+48 q^2+84 q+78\right) , \\
g_4(q, p) &= g_2(q, p).
\eea{gsPol}
The solution for the OPE data completely fixes the Wilson coefficients $\xi^{(1)}_{a,b}(p)$. At the next order in $\frac{1}{\sqrt{\lambda}}$ this procedure does not allow us to fix all the undetermined functions.

\section{Data and checks}\label{sec:DataAndChecks}

\subsection{OPE data}

We can determine all of the averages of the OPE data for each $\delta$ and $\ell$,
either by computing the residues of the poles of the world-sheet integral \eqref{A_to_B} and comparing with the residues computed in section \ref{sec:PolesResidues}, or by equating \eqref{taus} and \eqref{f2s} with \eqref{T2F2}. For example, for the leading Regge trajectory we find
\bea
\langle f_{s, 0} \tau_{s, 2} \rangle_{\delta, 2 \delta-2} ={}& \frac{r_0(\delta)}{2 \delta^{\frac{3}{2}}}(3 \delta^2 -\delta + 2), \\
\langle f_{t, 0} \tau_{t, 2} \rangle_{\delta, 2 \delta-2} ={}& \frac{r_0(\delta)}{2 \delta^{\frac{3}{2}}}(3 \delta^2 -\delta + \frac{p^2}{2}), \\
\langle f_{s, 2} \rangle_{\delta, 2 \delta-2} ={}& \frac{r_0(\delta)}{\delta^2} \Big( \frac{7 \delta ^2}{2}-\frac{7 \delta }{12}-\frac{p^3}{3}+\left(2 \delta ^2-\delta +\frac{1}{2}\right) p^2 \\
&+\frac{1}{6} \left(24 \delta ^2+3 \delta -1\right) p+\delta ^3 \left(2 \zeta (3)-\frac{7}{6}\right)-\frac{35}{32}\Big),  \\
\langle f_{t, 2} \rangle_{\delta, 2 \delta-2} ={}& \frac{r_0(\delta)}{\delta^2} \Big(\frac{7 \delta ^2}{2}+\frac{17 \delta }{12}+\frac{p^4}{8}+\left(\delta -\frac{13}{12}\right) p^3+\left(2 \delta ^2-\frac{7 \delta }{2}+\frac{23}{8}\right) p^2 \\
&+\frac{1}{6} \left(24 \delta ^2+3 \delta -28\right) p+\delta ^3 \left(2 \zeta (3)-\frac{7}{6}\right)+\frac{77}{32} \Big),
\eea{ResultsLeadingReggeTrajectory}
where
\begin{align}
r_n(\delta) = \frac{4^{2-2 \delta } \delta ^{2 \delta -2 n-1} (2 \delta -2 n-1)}{\Gamma (\delta ) \Gamma \left(\delta -\left\lfloor \frac{n}{2}\right\rfloor \right)}\,.
\end{align}
Using that the operators on the leading Regge trajectory are non-degenerate,
we can extract the twists for these operators with SU$(4)_R$ representation $[0,p-2,0]$ from the second equation in \eqref{ResultsLeadingReggeTrajectory} 
\bea
\tau^{[0,p-2,0]}\left(\tfrac{\ell}{2}+1,\ell\right) ={}& \sqrt{2 (\ell+2)} \lambda^\frac14 -\ell-2
+ \frac{3 \ell^2+10 \ell+2p^2 + 8}{4  \sqrt{2(\ell+2)}} \lambda^{-\frac14}
 + O(\lambda^{-\frac34})\,.
\eea{tau_leading_regge}
This is in agreement with \cite{Gromov:2011de}.\footnote{With the respect to the conventions of (2.28) of \cite{Gromov:2011de}, we need to identify $S=\ell+2$ and remember the shift $J=(p-2)+2$.} A few of the anomalous dimensions for $p>2$ have also been recomputed recently in \cite{Gromov:2023hzc} using the quantum spectral curve, namely
\bea
\tau_2^{[0,1,0]}(1,0) &= \frac{13}{4}\,, \quad
&\tau_2^{[0,2,0]}(1,0) &= 5\,, \qquad
\tau_2^{[0,3,0]}(1,0) = \frac{29}{4}\,,\\
\tau_2^{[0,1,0]}(2,2) &= \frac{29}{4\sqrt{2}}\,,\quad
&\tau_2^{[0,2,0]}(2,2) &= \frac{9}{\sqrt{2}}\,,
\eea{qsc_results}
also in perfect agreement with \eqref{tau_leading_regge}.

At the order we are considering, operators with odd spin only contribute to the quantity $\langle f_{t, 2} \rangle_{\delta, \ell}$, which is the leading contribution to the OPE coefficients for these operators. For the first few odd spin Regge trajectories our results for this quantity are
\begin{align}
\langle f_{t, 2} \rangle_{\delta\geq 2, 2 \delta-3} ={}&
 - \frac{r_{\frac12}(\delta)}{\delta } (\delta -1) \left(p^2-4\right) \,,\nonumber\\
\langle f_{t, 2} \rangle_{\delta\geq 3, 2 \delta-5} ={}&
 -\frac{2 r_{\frac32}(\delta)}{3}  (\delta -2) (\delta -1) (\delta +4) \left(p^2-4\right) \,, \label{f2t_odd}\\
\langle f_{t, 2} \rangle_{\delta\geq 4, 2 \delta-7} ={}&
 -\frac{r_{\frac52}(\delta)}{45}  (\delta -3) \left(10 \delta ^4+73 \delta ^3+128 \delta ^2-352 \delta -192\right) \left(p^2-4\right) \,.\nonumber
\end{align}
Anomalous dimensions for operators with odd spin appear only at higher orders in the $1/\sqrt{\lambda}$ expansion, i.e.\ in $A^{(k>1)}(S,T)$. In terms of degeneracies, the leading odd spin Regge trajectory with $\ell=2\delta-3$ is particularly simple: it was shown in \cite{Alday:2023flc} that all the operators on this trajectory have degeneracy 2.

\subsection{Wilson coefficients}
\label{app:literature}

The Wilson coefficients that appear in the low energy expansion \eqref{LEE} can be computed either from the world-sheet correlator \eqref{worldsheet_correlator}, as explained in \cite{Alday:2023jdk} (which uses the methods of \cite{Zerbini:blah}), or by plugging the solutions (\ref{taus}-\ref{f2s}) into the dispersive sum rule \eqref{EquationSecondOrder} and performing the sums, following \cite{Alday:2022xwz}. The first few terms in the expansion are given by
\begin{align}
{}&A^{(1)}(S,T) = - \frac{p}{S^2 T^2 U^2} \left(\frac{p}{6} S^2 + \frac{p-3}{3} T U \right)
+  p(p-2)  S \zeta (5) \label{LEE1}\\
&+\frac{\zeta (3)^2}{3}   \left(\left(p^2-6 p-36\right) S^2+2 \left(p^2+18\right) T U\right)
+\zeta (7) \left(2 p(p-2)  S^3- \left(2 p^2+2 p+\tfrac{441}{8}\right) S T U\right)\nonumber\\
&+\frac{2\zeta (3) \zeta (5)}{3}   \left(\left(p^2-10 p-81\right) S^4+2 \left(2 p^2+4
   p+81\right) S^2 T U-\left(2 p^2+4 p+81\right) T^2 U^2\right)+\ldots\nonumber\,.
\end{align}
Some values of the Wilson coefficients $\xi_{a, b}^{(1)}(p)$ had been computed before in the literature. Below we perform a detailed comparison of our results with all available results in the literature and we find full agreement.
In \cite{Binder:2019jwn} the Mellin amplitude was computed to order $\lambda^{-\frac{5}{2}}$ using localisation methods. We find full agreement when comparing their results to our prediction for $\xi_{a, 0}^{(1)}(p)$ for $a=0, 1$, which is
\begin{align}
\xi_{0, 0}^{(1)}(p) &=0\,, \\
\xi_{1, 0}^{(1)}(p) &= \frac{1}{4} (p-2) p \zeta (5)\,.
\end{align}
In \cite{Abl:2020dbx} an algorithm to compute the $p$-dependence of Wilson coefficients in a generic correlator $\langle \cO_{p_1} \cO_{p_2} \cO_{p_3} \cO_{p_4} \rangle$ was proposed. This algorithm is based on considering an effective field theory in full $AdS_5 \times S_5$. With this algorithm it is possible to compute the $p$-dependence of Wilson coefficients up to a finite number of undetermined parameters.
Take for example the Wilson coefficient $\xi_{2, 0}^{(1)}(p)$, which enters at $O(\lambda^{-3})$. \cite{Abl:2020dbx} predicts
\begin{align}\label{xi20abl}
\xi_{2, 0}^{(1)}(p) = \frac{1}{24} \zeta (3)^2 (3 \tilde{B}_1-24 \tilde{E}_0+(p-6) p-20),
\end{align}
where $\tilde{B}_1$ and $\tilde{E}_0$ are undetermined rational numbers.\footnote{More specifically, $B_1 = \tilde{B}_1 \zeta^2(3)$ and $E_0 = \tilde{E}_0 \zeta^2(3)$ are defined in equation $(84)$ of \cite{Abl:2020dbx}.} Our own computation gives
\begin{align}\label{xi20us}
\xi_{2, 0}^{(1)}(p) = \frac{1}{24} ((p-6) p-36) \zeta (3)^2\,.
\end{align} 
We see that (\ref{xi20abl}) and (\ref{xi20us}) are fully compatible and together they imply $3 \tilde{B}_1 - 24 \tilde{E}_0 + 16 =0 $.
Combining  this approach with our results and extending the localisation constraints in~\cite{Binder:2019jwn} at order $\lambda^{-3}$ we were able to fix the Mellin amplitude up to an unknown number $\beta$
\begin{align} \begin{aligned}
&M(s_1, s_2)\Big|_{\lambda^{-3}}=\frac{  \zeta_3^2(p)_4}{8 \Gamma (p-1)} \Big( (p+4)_3 \, s_1 s_2 s_3+\frac{(p+4)_2}{3}(p^2-6p-36)\,  s_1^2\\ &+\frac{2(p+4)_2}{3} (p^2+18)s_2 s_3
 +\frac{(p+4)}{3}(p-2)  ((p^3-7p^2-24p+36)+\frac{3}{2}\beta(p+2)) s_1\\
 &+\frac{2(p-12)}{27}(p^5-3p^4+2p^3+138p^2+288p+216)+\frac{\beta}{3}(p-2)^2p(p+2)\Big)\, .
 \end{aligned}
\end{align}

\section{Conclusions}\label{sec:Conclusions}

In this paper we determined the first curvature correction to the AdS Virasoro-Shapiro amplitude with external KK modes, for the correlator $\< \cO_2 \cO_2 \cO_p \cO_p \>$,
generalising the $\< \cO_2 \cO_2 \cO_2 \cO_2 \>$ results of \cite{Alday:2022uxp, Alday:2022xwz, Alday:2023jdk}.
We expect that higher curvature corrections can be obtained in an analogous way, as demonstrated in \cite{Alday:2023mvu}, where the second curvature correction for $\< \cO_2 \cO_2 \cO_2 \cO_2 \>$ was computed.

We also used the opportunity to recapitulate and contrast the two methods that were previously used to solve the problem at hand.
The method introduced in \cite{Alday:2023mvu} and used in section \ref{sec:Worldsheet} assumes the existence of a world-sheet integral with insertions of single-valued multiple polylogs,
which is suggested by the fact that world-sheet non-linear sigma models for AdS expanded around flat space receive contributions from flat space string amplitudes with insertions of additional soft gravitons \cite{Alday:2023jdk,Alday:2023mvu}. The ansatz is then fixed essentially by consistency with the OPE.

The method used in section \ref{sec:Dispersion} was introduced earlier, in \cite{Alday:2022xwz}, and involves an ansatz for certain combinations of OPE data in terms of Euler-Zagier sums.
This ansatz was inspired by the structure of the dispersive sum rules and their leading solution which can be inferred from flat space.
Of course, both methods lead to the same results.

An obvious future direction in the same line as the present paper would be to consider the correlator with generic half-BPS operators $\< \cO_{p_1} \cO_{p_2} \cO_{p_3} \cO_{p_4} \>$.
This would give access to operators in even more general $\text{SU}(4)_R$ representations, see \eqref{eq:exchangedRG}. 

Another generalisation would be to relax the assumption $p \ll \lambda^{\frac14}$.
As discussed in \cite{Alday:2023flc} (see also \cite{Aprile:2020luw}), in the regime $p \ll \lambda^{\frac14}$ all operators that are not KK modes of the $\text{SO}(5)$ singlet are suppressed by powers of $\lambda^{-\frac14}$. Studying correlators in the regime $p \sim \lambda^{\frac14}$ would make it easier to access those operators, which can also be characterised as having momentum in the $S^5$ directions.

\section*{Acknowledgements} 

We thank Fernando Alday,  Joseph Minahan, Erik Panzer and Oliver Schnetz for useful discussions.  GF thanks the Oxford Mathematical Institute for hospitality during the early stages of this work and Agnese Bissi, Parijat Dey,  Simon Ekhammar and Dima Volin for valuable discussions. 
The work of GF is supported by Knut and Alice Wallenberg Foundation under grant KAW 2016.0129 and by VR grant 2018-04438.
The work of TH is supported by the European Research Council (ERC) under the European Union's Horizon 2020 research and innovation programme (grant agreement No 787185). JS is supported by the STFC grant ST/T000864/1. 

\appendix

\section{$\mathcal{N}=4$ stress tensor four point function}
\label{app:BasicsAppendix}

We normalise two point functions of half-BPS operators $\mathcal{O}_{p}(x, y)$ as
\begin{align}
\langle \mathcal{O}_{p}(x_1, y_1) \mathcal{O}_{p}(x_2, y_2) \rangle = g_{12}^p
\end{align}
with $g_{ij}\equiv \frac{y_{ij}}{x_{ij}^2}$, where $y_{ij}=y_i \cdot y_j$ and $x_{ij}=x_i-x_j$. The correlator of four generic half-BPS operators $\mathcal{O}_{p_i}$ can be written as~\cite{Dolan:2003hv}
\begin{align}
\langle\mathcal{O}_{p_1}(x_1, y_1) \mathcal{O}_{p_2}(x_2, y_2) \mathcal{O}_{p_3}(x_3, y_3) \mathcal{O}_{p_4} (x_4, y_4) \rangle&=\mathcal{K}_{\{p_i\}} (x_i, y_i)\mathcal{G}_{\{ p_i\}} (z, \zb; \alpha, \bar{\alpha})\, ,\\
\mathcal{K}_{\{p_i\}} (x_i, y_i)&=g_{12}^{\frac{p_1+p_2}{2}}g_{34}^{\frac{p_3+p_4}{2}}\left( \frac{g_{24}}{g_{14}} \right)^{\frac{p_{21}}{2}} \left( \frac{g_{13}}{g_{14}} \right)^{\frac{p_{34}}{2}},
\end{align}
where $p_{ij}=p_i-p_j$.
We have introduced the cross-ratios
\begin{align}
U&=\frac{x_{12}^2 x_{34}^2}{x_{13}^2 x_{24}^2}=z \zb\, , \qquad \qquad V=\frac{x_{14}^2 x_{23}^2}{x_{13}^2 x_{24}^2}=(1-z)(1-\zb)\, ,\\
\frac{1}{\sigma}&=\frac{y_{12}y_{34}}{y_{13}y_{24}}=\alpha \bar{\alpha}\, , \qquad \qquad \frac{\tau}{\sigma}=\frac{y_{23}y_{14}}{y_{13}y_{24}}=(1-\alpha)(1-\bar{\alpha})\,  . 
\end{align}
If we consider the $s$-channel \textit{super}conformal block decomposition, $\mathcal{G}_{\{p_i\}}$ contains all the SU$(4)_R$ representations in $[0, p_1-2,0]\otimes [0, p_2-2, 0] \cap [0, p_3-2,0]\otimes [0, p_4-2, 0]$. These can be labelled as
\begin{align}\label{eq:exchangedRG}
[n-m, 2m+s, n-m]\,, \ n=0,  \cdots P=\mathrm{min}\left( p_i, \frac{p_i+p_j+p_k-p_l}{2}\right)-2\,, \  m=0,  \cdots,  n\,,
\end{align}
with $s= \mathrm{max}(|p_{12}|, |p_{34}|)$.

We decompose $\mathcal{G}_{\{p_i\}}$ into its free-theory value plus the  \textit{reduced correlator} $\mathcal{T}_{\{p_i\}}$
\begin{align}\label{eq:reducedCorrelator}
\mathcal{G}_{\{p_i\}}(z, \zb; \alpha, \bar{\alpha})=\mathcal{G}_{\{p_i\}}^{\mathrm{free}}(z, \zb; \alpha, \bar{\alpha})+\frac{(z-\alpha)(z-\bar{\alpha})(\zb-\alpha)(\zb-\bar{\alpha})}{(z\zb)^2(\alpha \bar{\alpha})^2}\mathcal{T}_{\{p_i\}}(z, \zb; \alpha, \bar{\alpha})\,. 
\end{align}
In the case under analysis, namely the one with two $\mathcal{O}_2$'s and two $\mathcal{O}_p$'s, the free correlator takes the form
\begin{align} \label{eq:Gfree}
\cG^{\mathrm{free}}_{\{22pp\}}=
g_{12}^2 g_{34}^p
+ \frac{p}{2 c} \left(g_{12} g_{14} g_{23} g_{34}^{p-1}+g_{12} g_{13} g_{24}
   g_{34}^{p-1}\right)+\frac{p(p-1)}{2c} g_{13} g_{14} g_{23} g_{24} g_{34}^{p-2}\,.
\end{align}
Only one representation is exchanged in the OPE decomposition and  $\mathcal{T}_{\{p_i\}}(z, \zb;\alpha, \bar{\alpha})$ is just a known function of $\alpha, \bar{\alpha}$ times a function of the cross-ratios

\bea
\cT_{\{22pp\}} (z,\zb;\al,\ab) &= \cT_{\{22pp\}} (U,V)\,,\\
\cT_{\{p22p\} / \{2p2p\}} (z,\zb;\al,\ab) &= (\al \ab)^{\frac{2-p}{2}} \cT^\prime_{\{p22p\} / \{2p2p\}} (U,V)\,.
\eea{eq:Hshorthand}

By imposing superconformal Ward identities, it is possible to disentangle contributions from short, protected operators and long, unprotected ones in $\cT_{\{p_i\}}$ --- see \cite{Nirschl:2004pa, Dolan:2004iy,Dolan:2001tt,Caron-Huot:2018kta, Aprile:2017xsp} for details.  So we can write
\begin{align}
\cT_{\{p_i \}}=\cT_{\{p_i \}}^{\mathrm{short}}+\cT_{\{p_i \}}^{\mathrm{long}}\,, 
\end{align}
and each  part can be expanded in superconformal blocks.
For the correlators under analysis, the protected contribution reads
\begin{align}
\cT_{\{22pp\}}^\text{short} (z,\zb;\al,\ab) &
=-\sum_{\lambda \geq 1}^{\infty}A_{2[\lambda+1]}^{(s)} G^{(0,0)}_{\lambda+5,\lambda-1} (z,\zb)\,, \nonumber \\
\cT_{\{2p2p\} }^\text{short} (z,\zb;\al,\ab) & 
=- (\a \ab)^{\frac{2-p}{2}} \sum\limits_{\l \geq 1}^\infty A^{(t)}_{p[\l+1]} G^{(p-2,2-p)}_{\lambda+p+3,\lambda-1} (z,\zb)\,, \label{shortH}\\
\cT_{\{p22p\} }^\text{short} (z,\zb;\al,\ab) &
=- (\a \ab)^{\frac{2-p}{2}} \sum\limits_{\l \geq 1}^\infty (-1)^\lambda A^{(t)}_{p[\l+1]} G^{(2-p,2-p)}_{\lambda+p+3,\lambda-1} (z,\zb)\,,\nonumber
\end{align}
where we have introduced the conformal blocks
\begin{align}
G_{\Delta,\ell}^{(r,s)} (z,\bar{z}) &= \frac{z \bar{z}}{\bar{z}-z} \left[ k_{\frac{\Delta-\ell-2}{2}}^{r,s}(z) k_{\frac{\Delta+\ell}{2}}^{r,s}(\bar{z}) - k_{\frac{\Delta+\ell}{2}}^{r,s}(z) k_{\frac{\Delta-\ell-2}{2}}^{r,s}(\bar{z}) \right], \label{Eq:G-block}
\\
k_h^{r,s}(z) &= z^h \; _{2}F_1 \left(h+ \frac{r}{2}, h + \frac{s}{2}; 2h, z \right)\, .
\end{align}
The OPE coefficients appearing in~\eqref{shortH},  necessary to reproduce the supergravity correlator in the dispersion relation, are given by
\begin{align}\begin{aligned}
A^{(s)}_{2[\lambda]}&= \frac{p}{2 c} \frac{1 + (-1)^\l}{2} \, \frac{2  (\lambda!)^2}{(2 \lambda)!}\,,\\
A^{(t)}_{p[\l]}&= \frac{p}{2c} \frac{\Gamma (p+\lambda -1) \left( (p-1)\Gamma (\lambda +1)+ (-1)^{\lambda }(p-1)_{\lambda }\right)}{\Gamma (p+2 \lambda -1)}\,.
\end{aligned}
\end{align}
Finally, the long contribution can be expanded in superconformal blocks as
\begin{align}\nonumber
\cT_{\{22pp\}}^\text{long} (z,\zb,\al,\ab) &
= \sum\limits_{\De,\ell} C_s\,  G^{(0,0)}_{\De+4,\ell} (z,\zb)\,,\\
\cT_{\{2p2p\} }^\text{long} (z,\zb,\al,\ab)&
= (\a \ab)^{\frac{2-p}{2}} \sum\limits_{\De,\ell} C_t\, G^{(p-2,2-p)}_{\De+4,\ell} (z,\zb)\,,\label{eq:Tlong_H}\\
\cT_{\{p22p\} }^\text{long} (z,\zb,\al,\ab) &
= (\a \ab)^{\frac{2-p}{2}} \sum\limits_{\De,\ell} (-1)^\ell C_t\, G^{(2-p,2-p)}_{\De+4,\ell} (z,\zb)\,.\nonumber
\end{align}

\section{$\text{SUGRA}^{(k)}$}
\begin{align} \label{eq:sugraGeneric}
&\begin{aligned}
&\mathrm{SUGRA}^{(1 \leq k  \leq p)}=  \frac{2^{-k}\Gamma(1+p)}{\Gamma(1+p-k)} \frac{1}{(S T U)^{k+1}} \Bigg\lbrace \mleft(-\frac{p}{3}\mright)^k S^{2k}+\mleft(-\frac{2}{3}(p-3)\mright)^k (T U)^k \\
&\,  +\sum_{j=1}^{\lfloor \frac{k}{2}  \rfloor}\mleft(-\frac{p}{3} \mright)^j \mleft( -\frac{p-6}{3}\mright)^j \mleft(-\frac{2}{3}(p-3) \mright) ^{k-2j}S^{2j} (T U)^{k-j}+\!\!\!\!\!\sum_{j=\lfloor \frac{k}{2}\rfloor+1}^{k-1} \mleft(-\frac{p}{3} \mright)^j S^{2j} (T U)^{k-j} \times \\
&\, \times \sum_{i=0}^{k-j}\mleft( -\frac{p-6}{3}\mright)^{k-j-i} \mleft(-\frac{2}{3}(p-3) \mright) ^{i} \frac{j(i+2 j-k+1)_{-i-j+k-1}(-i-2 j+k+1)_i}{i! (-i-j+k)!}
\Bigg\rbrace\, , 
\end{aligned}\\
&\phantom{{}^{1\leq }}\mathrm{SUGRA}^{(k>p)}=0 \, .
\end{align}
\section{Mack polynomials}\label{app:Macks}
To reproduce the superconformal block expansion of $\cT(U,V)$ in~\eqref{eq:Tlong_H}, the Mellin amplitudes should have poles at  precise locations, whose residues can be shown to be given in terms of  Mack polynomials $Q_{\ell, m}^{\Delta_{12}, \Delta_{34}, \tau}(s)$~\cite{Mack:2009mi, Dey:2016mcs}
\bea
\mathcal{Q}_s(s_3; \tau, \ell, m) &= \kappa^{({4,4,p+2,p+2})}_{\ell, m, \tau+4}  Q_{\ell, m}^{0, 0, \tau+4} \left(s_3-2-\frac{p}{3}\right)\, ,\\
\mathcal{Q}_t(s_3; \tau, \ell, m) &= \kappa^{({p+2, 4, 4, p+2})}_{\ell, m, \tau+4}  (-1)^{\ell}  Q_{\ell, m}^{p -2, 2-p, \tau+4} \left(s_3 -\frac{4}{3}p\right)\, .
\eea{MacksTwoChannel}
We  obtain a closed expression for $Q_{\ell, m}$ starting from (3.47) of~\cite{Dolan:2011dv},
\begin{align}
P_{i(\Delta-d/2), \ell}(s,t)=\frac{\sqrt{\pi } 2^{-d-\ell +3}  \Gamma (d+\ell -2)}{\Gamma \left(\frac{d-1}{2}\right) \Gamma \left(\frac{d}{2}+\ell -1\right)} \alpha _{\ell }\, ,
\end{align}
where $P_{\nu, \ell}(s,t)$ is the Mellin transform for the conformal partial wave in the convention of~\cite{Costa:2012cb}. From this data, it is straightforward to obtain a compact expression for the Mack polynomials, by partially resumming the expression of $\alpha_\ell$ 
\begin{align}
Q_{\ell, m}^{\Delta_{12}, \Delta_{34}, \tau}(s)&=\frac{2^{\ell } \ell ! \Gamma (\ell +\tau -1)}{\Gamma (2 \ell +\tau -1)} \sum_{k=0}^\ell \sum_{n=0}^{\ell-k} (-m)_k \left(m+\frac{s+\tau}{2}\right)_n \mu(\ell, k,n, \Delta_{12}, \Delta_{34}, \tau)\,,\nonumber\\
\mu(\ell, k,n, \Delta_{12}, \Delta_{34}, \tau)&=\frac{(-1)^{k+n+1} (\ell +\tau -1)_n }{k! n! (-k-n+\ell )!} \mleft(n+1+\frac{\Delta_{34}-\Delta_{12}}{2}\mright)_k \nonumber \\ 
&\times \quad\left(k+n-\frac{\Delta_{12}}{2}+\frac{\tau }{2}\right)_{-k-n+\ell }
   \left(k+n+\frac{\Delta_{34}}{2}+\frac{\tau }{2}\right)_{-k-n+\ell } \\ \nonumber
   &\times \, _4F_3\left( \begin{array}{cc}
-k,-1-n-\ell ,-1+\frac{\Delta_{12}}{2}+\frac{\tau
   }{2},-1-\frac{\Delta_{34}}{2}+\frac{\tau }{2}
   \\ \nonumber
 -\ell, \frac{\Delta_{12}}{2}-\frac{\Delta_{34}}{2}-k-n,-2+\tau
   \end{array} ; 1\right)\,,\nonumber
\end{align}
and
\begin{align}
\kappa_{\ell, m, \tau}^{(p_1, p_2, p_3, p_4)}&=-\frac{2^{1-\ell } (\ell +\tau -1)_{\ell } \Gamma (2 \ell +\tau ) \Gamma \mleft(-\frac{p_1 - p_2}{2}+\ell +\frac{\tau }{2}\mright)}{m! \left(-1+\ell +\tau\right)_m \Gamma \mleft(-m+\frac{p_1}{2}+\frac{p_2}{2}-\frac{\tau }{2}\mright) \Gamma
   \mleft(-m+\frac{p_3}{2}+\frac{p_4}{2}-\frac{\tau }{2}\mright)}  \\
   \nonumber &\times \Gamma \mleft(\frac{p_1 - p_2}{2}+\ell
   +\frac{\tau }{2}\mright) \Gamma \mleft(-\frac{p_3 - p_4}{2}+\ell +\frac{\tau }{2}\mright) \Gamma
   \mleft(\frac{p_3 - p_4}{2}+\ell +\frac{\tau }{2}\mright).\, \nonumber
\end{align}
We checked that these polynomials satisfy the expected recursion relations~\cite{Costa:2012cb}
\begin{align}
(\mathcal{D}_s-\lambda_\ell) Q_{\ell, m}(s)={}&4m\left(\frac{d}{2}-\tau-\ell-m\right)(2 Q_{\ell, m}(s)-Q_{\ell, m-1}(s+2)-Q_{\ell, m-1}(s))\,,\nonumber\\
 \mathcal{D}_s Q_{\ell, m}(s)={}&((2 m+s+\tau ) Q_{\ell,m}(s+2)-2 s Q_{\ell,m}(s)) (\Delta_{12}-\Delta_{34}+2 m+s+\tau )\,,\nonumber\\
 &+(\Delta_{12}+s) (s-\Delta_{34}) Q_{\ell,m}(s-2)\, , \label{mack_recursion}\\
 \lambda_\ell={}&(\Delta_{12}+2 m+\tau ) (-\Delta_{34}+2 m+\tau )+4 \ell  (2 m+\tau -1)+4 \ell ^2 \,.
\nonumber
\end{align}
In particular the case $m=0$ takes a very simple form
\begin{align}
Q_{\ell,  0}^{(\Delta_{12},\Delta_{34})}(s)=- 2^\ell\frac{\left(\frac{\tau }{2}-\frac{\Delta_{12}}{2}\right)_{\ell } \left(\frac{\Delta_{34}}{2}+\frac{\tau }{2}\right)_{\ell }}{(\ell +\tau -1)_{\ell }} \, _3F_2\left(\begin{array}{cc}-\ell ,\frac{s}{2}+\frac{\tau }{2},\tau +\ell -1\\ \nonumber
\frac{\tau }{2}-\frac{\Delta_{12}}{2},\frac{\Delta_{34}}{2}+\frac{\tau }{2}\end{array};1\right)\, .
\end{align}

\section{Large twist sums}\label{app:LargeTwist}

Given the dispersion relation
\bea
 M(s_1, s_2)= \sum_{\O_s} \sum_{m=0}^{\infty} C_{s} \frac{\mathcal{Q}_s(s_3; \tau_s, \ell, m) }{s_1 - \tau_s-2m + \frac{2}{3}p} + \sum_{\O_t} \sum_{m=0}^{\infty} C_{t} \frac{\mathcal{Q}_t(s_3;  \tau_t, \ell, m) }{s_2- \tau_t-2m + \frac{2}{3}p} 
\eea{TwoChannelDispApp}
and given that the operators exchanged have twist $\tau \sim \lambda^{1/4}$, we ask how do the 
OPE coefficients $C_s$, $C_t$ depend on $\lambda$ in such a way as to reproduce the first stringy correction
\begin{align}\label{FirstStringy}
\frac{p (p+1) (p+2) (p+3)}{\lambda ^{3/2} \Gamma (p-1)} \times \zeta(3).
\end{align}

At large $\tau$ the sum in $m$ in (\ref{TwoChannelDispApp}) is dominated by terms of order $\tau^2$. The large $\tau$ expansion of (\ref{TwoChannelDispApp}) can be obtained by setting $m \rightarrow x \tau^2$ and replacing $\sum_{m=0}^{\infty} \rightarrow \int dx \tau^2$. In this limit
\bea
\lim_{\tau_s \rightarrow \infty} \int_0^{\infty} dx \tau_s^2 \frac{
\mathcal{Q}_s(s_3; \tau_s, \ell, x \tau_s^2) }{s_1- \tau_s-2 x \tau_s^2 + \frac{2}{3}p} &=\frac{(\ell+1) (-1)^{-p}  \Gamma (p+4) \sin ^2\left(\frac{\pi  \tau_s }{2}\right) 2^{2 \ell+2 p+2 \tau_s +13}}{\pi ^3 \tau_s ^{2 (p+4)}}, \\
\lim_{\tau_t \rightarrow \infty} \int_0^{\infty} dx \tau_t^2 \frac{\mathcal{Q}_t(s_3;  \tau_t, \ell, x \tau_t^2) }{s_2- \tau_t-2 x \tau_t^2+ \frac{2}{3}p}  &= \frac{(-1)^{\ell} (\ell+1)  \Gamma (p+4) 2^{2 \ell+2 p+2 \tau_t +13} \sin ^2\left(\frac{1}{2} \pi  (p-\tau_t )\right)}{\pi ^3 \tau_t ^{2 (p+4)}}.
\eea{LargeTauMacks}
Inserting (\ref{LargeTauMacks}) and (\ref{FirstStringy}) into (\ref{TwoChannelDispApp}) we derive (\ref{OPEStringy}).

\section{Ambiguities of the world-sheet integrand}
\label{app:ambiguities}

The ambiguous terms in the world-sheet correlator \eqref{worldsheet_correlator} are given by
\begin{align}
\hat{r}^{(1)s}_{1,1} ={}& \left(0,0,0,\tfrac{S^2+4 S T+T^2}{(S+T)^2}\right) \,, \quad 
\hat{r}^{(1)a}_{1,1} = \left(0,0,0\right)\,,\quad
\hat{r}^{(1)s}_{2,1} = \left(0,0,0,0\right)\,,\quad
\hat{r}^{(1)a}_{2,1} = \left(0,0,0\right)\,,\nonumber\\
\hat{r}^{(1)s}_{1,2} ={}& \left(-\tfrac{5 S^2+8 S T+5 T^2}{12 (S+T)^2},\tfrac{S^2+S T+T^2}{3 (S+T)^2},-\tfrac{5 S^2+8 S T+5 T^2}{12 (S+T)^2},4\right) \,, \nonumber\\
\hat{r}^{(1)a}_{1,2} ={}& \left(-\tfrac{5 (S-T)}{12 (S+T)},\tfrac{S-T}{3 (S+T)},\tfrac{S-T}{4 (S+T)}\right)\,,\quad
\hat{r}^{(1)s}_{2,2} = \left(0,0,0,0\right)\,, \quad
\hat{r}^{(1)a}_{2,2} = \left(0,0,0\right)\,,\nonumber\\
\hat{r}^{(1)s}_{1,3} ={}& \left(0,0,0,0\right) \,, \quad 
\hat{r}^{(1)a}_{1,3} = \left(0,0,0\right)\,,\quad
\hat{r}^{(1)s}_{2,3} = \left(0,0,0,\tfrac{2 \left(S^2+2 S T\right)}{(S+T)^2}\right)\,,\quad
\hat{r}^{(1)a}_{2,3} = \left(0,0,0\right)\,,\nonumber\\
\hat{r}^{(1)s}_{1,4} ={}& \left(0,0,0,1\right) \,, \quad 
\hat{r}^{(1)a}_{1,4} = \left(0,0,0\right)\,,\quad
\hat{r}^{(1)s}_{2,4} = \left(0,0,0,-\tfrac{2 S+T}{S+T}\right)\,,\quad
\hat{r}^{(1)a}_{2,4} = \left(0,0,0\right)\,,\nonumber\\
\hat{r}^{(1)s}_{1,5} ={}& \left(1,0,1,0\right) \,, \ 
\hat{r}^{(1)a}_{1,5} = \tfrac{S-T}{S+T} \left(1,-2,-1\right)\,,\ 
\hat{r}^{(1)a}_{2,5} = \left(-\tfrac{T^2-2 S^2}{4 (S+T)^2},-\tfrac{2 S^2+9 S T+8 T^2}{2 (S+T)^2},-\tfrac{(S+2 T)^2}{4 (S+T)^2}\right)\,,\nonumber\\
\hat{r}^{(1)s}_{2,5} ={}& \left(\tfrac{2 S^2+T^2}{4 (S+T)^2},-\tfrac{2 S^2+3 S T+4 T^2}{2 (S+T)^2},-\tfrac{S^2-4 T^2}{4 (S+T)^2},\tfrac{4 \left(4 S^2+9 S T-4 T^2\right)}{(S+T)^2}\right)\,,\nonumber\\
\hat{r}^{(1)s}_{1,6} ={}& \left(0,1,0,0\right) \,, \quad 
\hat{r}^{(1)a}_{1,6} = \left(0,0,0\right)\,,\quad
\hat{r}^{(1)a}_{2,6} = \left(\tfrac{6 S^2+4 S
   T+T^2}{8 (S+T)^2},-\tfrac{6 S^2+7 S T+4 T^2}{4 (S+T)^2},-\tfrac{3 S^2}{8 (S+T)^2}\right)\,,\nonumber\\
\hat{r}^{(1)s}_{2,6} ={}& \left(\tfrac{6 S^2+4 S T-T^2}{8 (S+T)^2},-\tfrac{6 S^2+13 S T+8 T^2}{4 (S+T)^2},\tfrac{-3 S^2+4 S T+8 T^2}{8 (S+T)^2},\tfrac{8 S^2+6 S T-8 T^2}{(S+T)^2}\right)\,.
\label{ambiguities}
\end{align}

\section{Crossing-symmetric dispersion relation}\label{app:ImprovedDisp}

Our Mellin amplitude has the symmetry $M(s_1, s_2) = M(s_1, s_3)$. Following \cite{Gopakumar:2021dvg, Sinha:2020win}, it is useful to construct a dispersion relation where this symmetry is manifest, since that will lead to simpler dispersive sum rules.\\
We will use the variables $s_1$ and $r$, where $r= \frac{s_2 s_3}{s_1}$. In terms of these variables we have
\beq
s'_2(s_1,r) = \frac{1}{2} \left(-s_1 + \sqrt{s_1(s_1-4 r)}\right)\,, \qquad
s'_3(s_1,r) = \frac{1}{2} \left(-s_1 - \sqrt{s_1(s_1-4 r)}\right)\,,
\label{s23_to_s1r}
\eeq
and we will consider
\begin{align}
\widetilde{M} (s_1, r)= M(s_1, s'_2(s_1,r)).
\end{align}
Although $s'_2(s_1,r)$ has a branch cut from $s_1=0$ to $s_1=4 r$, $\widetilde{M}(s_1, r)$ does not have a branch cut because of the symmetry $M(s_1, s'_2(s_1,r)) = M(s_1, s'_3(s_1,r))$.
In the new variables the bound on chaos (\ref{bound chaos}) becomes
\beq
\widetilde{M}(s_1, r) = O(s_1^{-2})\,, \text{ for } |s_1|\to \infty \text{ with } \text{Re}(r)> -\frac{p}{3}\,.
\eeq
Thus, keeping $r$ fixed and with $\text{Re}(r)> -\frac{p}{3}$ we can write a dispersion relation in the new variables
\beq
\widetilde{M}(s_1, r) = \oint_{s_1} \frac{ds_1'}{2 \pi i}  \frac{\widetilde{M}(s_1', r )}{(s_1'-s_1)}\,.
\label{dr_start}
\eeq
In the $s_1$ plane $\widetilde{M}(s_1, r)$ has poles at $s_1 = \tau_s^m$ and $s_1=-\frac{(\tau_t^m)^2}{\tau^m_t+r}$, where $\tau^m_s = \tau_s + 2m - \frac{2}{3}p$ and $\tau_t^m = \tau_t + 2m - \frac{2}{3}p$. This leads to
\bea
\widetilde{M}(s_1, r) = \sum_{\O_s} \sum_{m=0}^{\infty} C_{s} \frac{\tilde{\mathcal{Q}}_s(s'_2(\tau_s^m,r)-2-\frac{p}{3})}{s_1-\tau_s^{m}}
- \sum_{\O_t} \sum_{m=0}^{\infty}  C_t \frac{\tau_t^m(2r+\tau_t^m)}{(r+\tau_t^m)^2}\frac{\tilde{\mathcal{Q}}_t(-2-\frac{p}{3} - \frac{(\tau_t^m)^2}{r+\tau_t^m})}{s_1 + \frac{(\tau_t^m)^2}{r+\tau_t^m}},
\eea{dispersion}
where 
\bea
\tilde{\mathcal{Q}}_s(x) &\equiv \kappa^{({4,4,p+2,p+2})}_{\ell, m, \tau+4}  Q_{\ell, m}^{0, 0, \tau+4} (x)\,, \\
\tilde{\mathcal{Q}}_t(x) &\equiv \kappa^{({p+2, 4, 4,p+2})}_{\ell, m, \tau+4}  Q_{\ell, m}^{2-p, 2-p, \tau+4} (x)\,.
\eea{Macktilde}
It is important to have formulas that connect OPE data with the low energy expansion of $\widetilde{M}(s_1, r)$. Expanding \eqref{dispersion} as
\begin{align}
\widetilde{M}(s_1, r) = \sum_{a,b=0}^{\infty} \xi_{a, b}s_1^{a} r^b\,,
\end{align}
the coefficients read
\begin{align}\label{MasterDisp}
\xi_{a, b} = \sum_{\O_s} \sum_{m=0}^{\infty} C_{s} \sum_{q=0}^b \mathfrak{U}^{\tau_s^m, \ s}_{a,b,q} \partial^q_{x} \mathfrak{Q}_s (x) |_{x=0} + \sum_{\O_t} \sum_{m=0}^{\infty} C_{t} \sum_{q=0}^b \mathfrak{U}^{\tau_t^m, \ t}_{a,b,q} \partial^q_{x} \mathfrak{Q}_t (x) |_{x=0}\,,
\end{align}
where
\begin{align}
\mathfrak{U}^{\tau, \, s}_{a,b,0} &= - \frac{\delta_{b,0}}{\tau^{a+1}}\,,\nonumber\\
\mathfrak{U}^{\tau, \ s}_{a,b,q>0} &=
\frac{(-1)^{q+1} \Gamma (2 b-q) }{\Gamma(b+1) \Gamma (q) \Gamma (b-q+1) \tau^{a+b+1-q}} \,,\nonumber\\
\mathfrak{U}^{\tau, \ t}_{a,b,0} &= \frac{(-1)^{-a-b} (a+b) (1-a)_{b-1} }{\Gamma (b+1)\tau^{a+b+1}}\,,\label{Ufrak}\\
\mathfrak{U}^{\tau, \ t}_{a,b,q>0} &= \frac{(-1)^{a-1} \Gamma (a) (a+b-q) \, _3F_2(q,q-b,-a-b+q+1;-a-b+q,a-b+q+1;1)}{\Gamma (q+1) \Gamma (b-q+1) \Gamma (a-b+q+1) \tau^{1+a+b-q}}\,,
\nonumber
\end{align}
and
\bea
\mathfrak{Q}_s (x) &\equiv \kappa^{({4,4,p+2,p+2})}_{\ell, m, \tau+4}  Q_{\ell, m}^{0, 0, \tau+4} (x -2 - \frac{p}{3}) \,, \\
\mathfrak{Q}_t (x) &\equiv  \kappa^{({p+2, 4, 4,p+2})}_{\ell, m, \tau+4} (-1)^{\ell}   Q_{\ell, m}^{p-2, 2-p, \tau+4} (-x-\frac{4}{3}p)\,.
\eea{WeirdQ}
Using formula (\ref{MasterDisp}) and the large twist expansions of Appendix~\ref{app:LargeTwist} we can derive the expressions relating the Wilson coefficients and the OPE data that we present in the main text.
Finally, the Wilson coefficients $\xi_{a, b}^{(0)}(p)$ can be read off from the flat space Virasoro-Shapiro amplitude through \cite{Alday:2019nin}
\small
\begin{align}\label{FlatSpace1}
&\frac{4p}{\Gamma(p-1)}\frac{1}{s_1 s_2 s_3} + M_{\text{flat}}(s_1,s_2) = \frac{4}{\Gamma(p) \Gamma(p-1) s_1 s_2 s_3} \int_0^{\infty} d\beta \, e^{-\beta} \, \beta^{p} \, A_{\text{flat}}(2 \beta s_1, 2 \beta s_2)\,,
\end{align}
\normalsize
where
\bea
A_{\text{flat}}(s_1, s_2) &= \exp\left(\sum_{k=1}^{\infty} \frac{2 \zeta(2k+1)}{2k+1} \left(\frac{1}{4 \sqrt{\lambda}}\right)^{2k+1}\left(s_1^{2k+1} + s_2^{2k+1} + s_3^{2k+1}\right) \right)\,,  \\
 M_{\text{flat}}(s_1,s_2) &= \sum_{a=0}^{\infty} \sum_{b=0}^{a} \frac{\Gamma (a+b+p+4)}{\Gamma (b+1) \Gamma(p) \Gamma(p-1)} s_1^{a-b} \big(s_2 s_3 \big)^b \lambda^{-\frac{3}{2}-\frac{a}{2}-\frac{b}{2}}  \xi_{a, b}^{(0)}(p) \,.
\eea{FlatSpace2}

\section{Details about dispersive sum rules}
\subsection{Formulas}\label{app:formulas}
The functions appearing in \eqref{EquationSecondOrder} are given by
\begin{align}
c_{s,a,b,q}^{(2,0)} &=  \frac{1}{96}\Big(-16 a^3-16 b^3-16 b^2 (p-3 q+6)+32 p^3-112 p^2-480 p+321  \nonumber\\
&-16 a^2 (3 b+p-3 q+6)-80 p^2 q-80 p q-168 q^2+228 q -48 p^2 q^2-144 p q^2  \nonumber\\
&+8 b \left(4 p^2+8 p q-10 p-3 q^2+36 q+2\right) \nonumber\\
&-8 a \left(6 b^2+4 b (p-3 q+6)-4 p^2-8 p q+10 p+3 q^2-36 q-2\right)  \Big)c_{s,a,b,q}^{(0)} \,,  \nonumber\\
c_{s,a,b,q}^{(2,1)} &=-\frac{1}{48} (q+1)\Big( -12 a^2-12 b^2-2 b (8 p-6 q+33)+32 p^2+56 p-15   \nonumber\\
                   &-2 a (12 b+8 p-6 q+33)+24 p^2 q+72 p q+84 q  \Big) c_{s,a,b,q}^{(0)}\,,  \nonumber\\
c_{t,a,b,q}^{(2,0)} &=  \frac{1}{96}\Big(-16 a^3-16 b^3-16 b^2 (p-3 q+6)+32 p^3-112 p^2-480 p+321 \nonumber\\
&-16 a^2 (3 b+p-3 q+6)-80 p^2 q-80 p q-168 q^2+228 q -48 p^2 q^2-144 p q^2  \nonumber\\
&+8 b \left(4 p^2+8 p q-10 p-3 q^2+36 q+2\right) -12 (p-2)^2 \left(2 a+2 b+p^2+6 p-2 q+17\right) \nonumber\\
&-8 a \left(6 b^2+4 b (p-3 q+6)-4 p^2-8 p q+10 p+3 q^2-36 q-2\right)  \Big)c_{t,a,b,q}^{(0)}  \,, \nonumber\\
c_{t,a,b,q}^{(2,1)} &=-\frac{1}{48} (q+1)\Big( -12 a^2-12 b^2-2 b (8 p-6 q+33)+32 p^2+56 p-15   \nonumber\\
                   &-2 a (12 b+8 p-6 q+33)+24 p^2 q+72 p q+84 q  \Big) c_{t,a,b,q}^{(0)}\,.               
\label{dump1}
\end{align}
\subsection{Dispersive sum rule for $b=0$}\label{app:bEquals0}

The simplest dispersive sum rule is the one for the flat space Wilson coefficient $\xi_{a, 0}^{(0)}(p)$. It leads to the equations
\begin{align}
\zeta(3+a) &= \frac{1}{2} \sum_{\delta=1}^{\infty}\frac{\sum_{\ell=0, 2}^{2 \delta-2} f_{s, 0}(\delta, \ell)}{\delta^{3+a}} + \frac{1}{2} \sum_{\delta=1}^{\infty}\frac{\sum_{\ell=0, 1}^{2 \delta-2} f_{t, 0}(\delta, \ell)}{\delta^{3+a}}, ~~~ a \in 2 \mathbb{N}_0\,, \\
0 &=  \sum_{\delta=1}^{\infty}\frac{\sum_{\ell=0, 2}^{2 \delta-2} f_{s, 0}(\delta, \ell)}{\delta^{3+a}} - \sum_{\delta=1}^{\infty}\frac{\sum_{\ell=0, 1}^{2 \delta-2} f_{t, 0}(\delta, \ell)}{\delta^{3+a}}, ~~~ a \in (2 \mathbb{N}_0 + 1)\,,
\end{align}
i.e.\ the first equation holds for $a$ even and the second for $a$ odd. This separation between $a$ even and odd has to do with the fact that in the flat space limit the Mellin amplitude only contains terms of type $(s_1^2 + s_2^2 + s_3^2)^{n_1} (s_1 s_2 s_3)^{n_2}$.
The equations for odd $a$ imply that 
\begin{align}
\sum_{\ell=0, 2}^{2 \delta-2} f_{s, 0}(\delta, \ell) =  \sum_{\ell=0, 1}^{2 \delta-2} f_{t, 0}(\delta, \ell).
\end{align}
As to the equations with even $a$, since $\zeta(3+a) = \sum_{\delta=1}^{\infty} \frac{1}{\delta^{3+a}}$ they imply that 
\begin{align}
\sum_{\ell=0, 2}^{2 \delta-2} f_{s, 0}(\delta, \ell) =  1.
\end{align}

\subsection{Example}\label{app:ExampleMaple}

In this Appendix we demonstrate how we solve (\ref{EquationSecondOrder}) for $b=0$. For $q=0$ the ansatz for the OPE data is 
\bea 
T_{s; q=0}^{(2)}(\delta)={}& d_1(p) Z(\delta-1) + d_2(p) \delta Z_1(\delta-1) +  d_3(p) \delta^2 Z_2(\delta-1)\,, \\
T_{t; q=0}^{(2)}(\delta)={}& d_4(p) Z(\delta-1) + d_5(p) \delta Z_1(\delta-1) +  d_6(p) \delta^2 Z_2(\delta-1)\,,  \\
F_{s; q=0}^{(2)}(\delta)={}& d_7(p) Z(\de-1) + d_8(p) \de Z_1(\de-1) +  d_9(p) \de^2 Z_2(\de-1) \\
&+ d_{10}(p) \de^3 Z_{3}(\de-1)+ d_{11}(p) \delta^3 \zeta(3)  Z(\delta-1),  \\
F_{t; q=0}^{(2)}(\delta)={}& d_{12}(p) Z(\de-1) + d_{13}(p) \de Z_1(\de-1) +  d_{14}(p) \de^2 Z_2(\de-1) \\
&+ d_{15}(p) \de^3 Z_{3}(\de-1)+ d_{16}(p) \delta^3 \zeta(3)  Z(\delta-1).  
\eea{qEquals0Ansatz}
$d_i(p)$ are $16$ functions of $p$ that we will fix by our procedure.\footnote{The terms proportional to $Z(\delta-1)$ are required, by contrast to the $q \geq 1$ case. The reason is that the functions in (\ref{T2F2}) vanish for $(\delta=1, \, q \geq 1)$, but do not vanish when $(\delta=1, \, q = 0)$.} 
We plug (\ref{qEquals0Ansatz}) into (\ref{EquationSecondOrder}) for $b=0$ and use the formula
\begin{align}
\zeta(s, s_1, s_2, \ldots) = \sum_{\delta=1}^{\infty} \frac{Z_{s_1, s_2, \ldots}(\delta-1)}{\delta^{s}}\,,
\end{align}
to express $\xi_{a, 0}^{(1)}(p)$ in terms of multiple zeta values and as a function of $d_i(p)$. Convergence of the $\delta$ sum implies that 
\begin{align}
d_{10}(p)+d_{11}(p)+d_{15}(p)+d_{16}(p)=0.
\end{align}
Finally, we impose that $\xi_{a, 0}^{(1)}(p)$ is a linear combination of single-valued multiple zeta values. We do so using the MAPLE program \textit{HyperlogProcedures} \cite{HyperlogProcedures}. It was enough to impose this for $a=0, \ldots, 6$. In this manner we fix $13$ out of the $16$ functions $d_i(p)$\footnote{After fixing the functions $d_i(p)$ in this way we checked that the resulting expression for $\xi_{a, 0}^{(1)}(p)$ is single-valued for $a=7, \ldots , 10$.}
\small
\bea
&d_2(p) = \frac{1}{4}, \, d_3(p) = 1, \, d_5(p) = \frac{1}{4}, \, d_6(p) = 1, d_7(p) = 2 d_1(p)+2 d_4(p)-d_{12}(p)+\frac{p^4}{8}-\frac{5 p^3}{12}+\frac{19 p^2}{8}+\frac{25 p}{6}\\
&+\frac{13}{16}, \, d_8(p)= 2 d_1(p)+2 d_4(p)+\frac{p^2}{2}+3 p-\frac{39}{8}, \, d_9(p)= 2, \, d_{10}(p)= -2, \, d_{13}(p)= 2 d_1+2 d_4+\frac{p^2}{2}+3 p\\
&-\frac{39}{8}, \, d_{14}(p)= 2, \, d_{15}(p)= -2, \, d_{16}(p)= 2  .
\eea{SolPards}
\normalsize
The $3$ functions $d_1(p)$, $d_4(p)$ and $d_{12}(p)$ are not fixed by this procedure. However, they are fixed by repeating this exercise for $b=0, 1, \ldots$\,.

\bibliographystyle{JHEP}
\bibliography{22pp}
\end{document}